\documentclass[twocolumn]{aastex631}

\newcommand\kms{km$\,$s$^{-1}$}
\newcommand\Msol{M$_{\odot}$}

\newcommand{\hi}{H\,{\sc i}}

\graphicspath{{./}{figures/}}

\begin{document}

\title{Star formation histories and gas content limits of three ultra-faint dwarfs on the periphery of M31}
\shorttitle{SFHs of three M31 UFDs}

\correspondingauthor{Michael G. Jones}
\email{jonesmg@arizona.edu}

\author[0000-0002-5434-4904]{Michael G. Jones}
\affiliation{Steward Observatory, University of Arizona, 933 North Cherry Avenue, Rm. N204, Tucson, AZ 85721-0065, USA}
\affiliation{IPAC, Mail Code 100-22, Caltech, 1200 E. California Blvd., Pasadena, CA 91125, USA}

\author[0000-0003-4102-380X]{David J. Sand}
\affiliation{Steward Observatory, University of Arizona, 933 North Cherry Avenue, Rm. N204, Tucson, AZ 85721-0065, USA}

\author[0000-0001-8354-7279]{Paul Bennet}
\affiliation{Space Telescope Science Institute, 3700 San Martin Drive, Baltimore, MD 21218, USA}

\author[0000-0002-1763-4128]{Denija Crnojevi\'{c}}
\affil{University of Tampa, 401 West Kennedy Boulevard, Tampa, FL 33606, USA}

\author[0000-0001-9775-9029]{Amandine~Doliva-Dolinsky}
\affil{University of Surrey, School of Mathematics and Physics, Guildford, GU2 7XH, UK}

\author[0000-0001-8245-779X]{Catherine E. Fielder}
\affiliation{Steward Observatory, University of Arizona, 933 North Cherry Avenue, Rm. N204, Tucson, AZ 85721-0065, USA}

\author[0000-0001-5368-3632]{Laura C. Hunter}
\affiliation{Department of Physics and Astronomy, Dartmouth College, 6127 Wilder Laboratory, Hanover, NH 03755, USA}

\author[0000-0001-8855-3635]{Ananthan Karunakaran}
\affiliation{Department of Astronomy \& Astrophysics, University of Toronto, Toronto, ON M5S 3H4, Canada}
\affiliation{Dunlap Institute for Astronomy and Astrophysics, University of Toronto, Toronto ON, M5S 3H4, Canada}

\author[0000-0001-9649-4815]{Bur\c{c}in Mutlu-Pakdil}
\affil{Department of Physics and Astronomy, Dartmouth College, Hanover, NH 03755, USA}

\author[0000-0002-8217-5626]{Deepthi S. Prabhu}
\affiliation{Steward Observatory, University of Arizona, 933 North Cherry Avenue, Rm. N204, Tucson, AZ 85721-0065, USA}

\author[0000-0002-0956-7949]{Kristine Spekkens}
\affiliation{Department of Physics, Engineering Physics and Astronomy, Queen’s University, Kingston, ON K7L 3N6, Canada}

\author[0000-0002-5177-727X]{Dennis Zaritsky}
\affiliation{Steward Observatory, University of Arizona, 933 North Cherry Avenue, Rm. N204, Tucson, AZ 85721-0065, USA}



\begin{abstract}
We present Hubble Space Telescope (HST) imaging of Pegasus~V and Pisces~VII, along with a re-analysis of the archival imaging of Pegasus~W, and Jansky Very Large Array (VLA) neutral gas (\hi) observations of all three. These three ultra-faint dwarfs (UFDs) are all within the Local Group in the approximate direction of M31.
The VLA observations place stringent upper limits on their \hi \ content, with all having $M_\mathrm{HI} < 10^4$~\Msol. 
As the red giant branches of these UFDs are sparsely populated, we determined distances from the HST photometry of horizontal branch (HB) stars in comparison to a fiducial HB population (from M92), with all three falling in the range 0.7-1~Mpc.
Using a new \texttt{Python}-based star formation history (SFH) fitting code (based on \texttt{StarFISH}), we derive SFHs of all three UFDs. As found previously, the best fit SFH for Pegasus~W includes significant star formation well beyond the end of reionization, while the SFHs calculated for Pegasus~V and Pisces~VII are consistent with them having quenched over 10~Gyr ago. These findings for the latter two objects indicate that, like those in the vicinity of the Milky Way, lower mass UFDs in the vicinity of M31 likely quenched at early times. 
\end{abstract}



\section{Introduction} \label{sec:intro}

Ultra-faint dwarfs \citep[UFDs; for a review see][]{Simon+2019} are the faintest class of galaxies known, with absolute magnitudes fainter than $M_\mathrm{V} = -7.7$, roughly corresponding to a total stellar mass $M_*\lesssim 10^5$~\Msol. Wide-area public imaging surveys such as the Sloan Digital Sky Survey \citep[SDSS;][]{York+2000}, the Dark Energy Survey \citep[DES;][]{DES_DR1}, the DECam Local Volume Exploration Survey \citep[DELVE;][]{Drlica-Wagner+2021}, and the DESI Legacy Imaging Surveys \citep{Dey+2019} have enabled large numbers of UFDs to be identified throughout the Local Group (LG) via searches for overdensities of resolved stars \citep[e.g.,][]{Willman+2005,Belokurov+2006,Irwin+2007,McConnachie+2008,Koposov+2015,Drlica-Wagner+2015,Cerny+2021}. This technique, in combination with extremely deep, targeted surveys of individual nearby Milky Way-like systems, has even resulted in the detection of a handful of UFDs beyond the LG \citep[e.g.,][]{Crnojevic+2016,Mutlu-Pakdil+2022}. Over the past few years there has been increasing interest in expanding beyond this approach in order to enable the detection of UFDs whose stellar bodies are not fully resolved into individual stars in existing wide-field surveys \citep[e.g.,][]{Sand+2022,Sand+2024,Martinez-Delgado+2022,Collins+2022,McQuinn+2023,Jones+2023b,Jones+2024,Li+2024}. These recent searches have used both machine learning and visual inspection (sometimes with the help of amateur astronomers) and have discovered a few new UFDs (and slightly more massive star-forming galaxies), including the first UFD known in isolation, well-beyond the edge of the LG \citep[Tucana~B;][]{Sand+2022}. 

Three of these recent discoveries are all in the approximate direction of M31: Pisces~VII \citep{Martinez-Delgado+2022,Collins+2024}, Pegasus~V \citep{Collins+2022}, and Pegasus~W \citep{McQuinn+2023}. 
All three objects were originally identified as partially resolved overdensities in the DESI Legacy Imaging Surveys, and then followed up with deep ground-based (for Pegasus~V and Pisces~VII) and Hubble Space Telescope (HST) imaging (for Pegasus~W). Distance estimates were determined using color-magnitude diagram (CMD) fitting techniques, revealing that all are in the UFD luminosity regime, and are significantly fainter than most of the UFDs near M31 that have been studied in detail to date \citep[e.g.][]{Savino+2023}.

Based on these distances, all three appear to be at the edge, or just beyond, the virial radius of M31. Pegasus~V and W may be on their first infall into the M31 system and might never have been within the halo of a more massive system. On the other hand, Pisces~VII is likely an M33 satellite \citep{Collins+2024}. Given the proximity of the M31 system it is possible to constrain the star formation histories (SFHs) of such objects with relatively short exposures with HST. This is of critical importance as UFDs at these luminosities are expected to have been quenched early on as a result of cosmic reionization \citep{Benson+2002,Bovill+2009,Simpson+2013,Wheeler+2015,Fitts+2017}. However, to date most of the small number of available SFHs of galaxies at these masses have been derived for UFD satellites of the Milky Way (MW) \citep[e.g.][]{Brown+2012,Brown+2014,Weisz+2014}, and a few just beyond its virial radius \citep{McQuinn+2024a}. As a large fraction of the MW satellites are thought to have first fallen into the system $\sim$10~Gyr ago \citep[e.g.][]{Weisz+2019,D'Souza+2021}, environmental processes could be an alternate explanation for the ancient quenching of these UFDs. Thus, Pisces~VII, Pegasus~V and Pegasus~W offer a key new opportunity to assess the cosmic reionization quenching hypothesis for UFDs in the vicinity of a different host galaxy that assembled its satellite system at a different time.

Remarkably, the existing HST observations of Pegasus~W \citep{McQuinn+2023} show tentative signs of recent ($\lesssim$1~Gyr) star formation (SF), suggesting that it may not have been permanently quenched by reionization and may even contain a gas reservoir capable of supporting future SF. Motivated by these findings, we followed up these UFDs with joint observations with the Jansky Very Large Array (VLA) and HST to search for signs of gas reservoirs via the neutral hydrogen 21~cm line emission (\hi) and to characterize their stellar populations and SFHs.

In the following section we describe these HST and VLA observations. In \S\ref{sec:struct_params} we derive structural parameters and distances for each object. In \S\ref{sec:SFHfitting} we describe our SFH fitting methodology in detail, and present our findings in \S\ref{sec:results}. These are discussed further in \S\ref{sec:dicussion}, before we present our conclusions in \S\ref{sec:conclusions}. 

\section{Observations} \label{sec:observations}

\subsection{HST imaging} \label{sec:HSTdata}

\begin{figure*}
    \centering
    \includegraphics[width=0.49\linewidth]{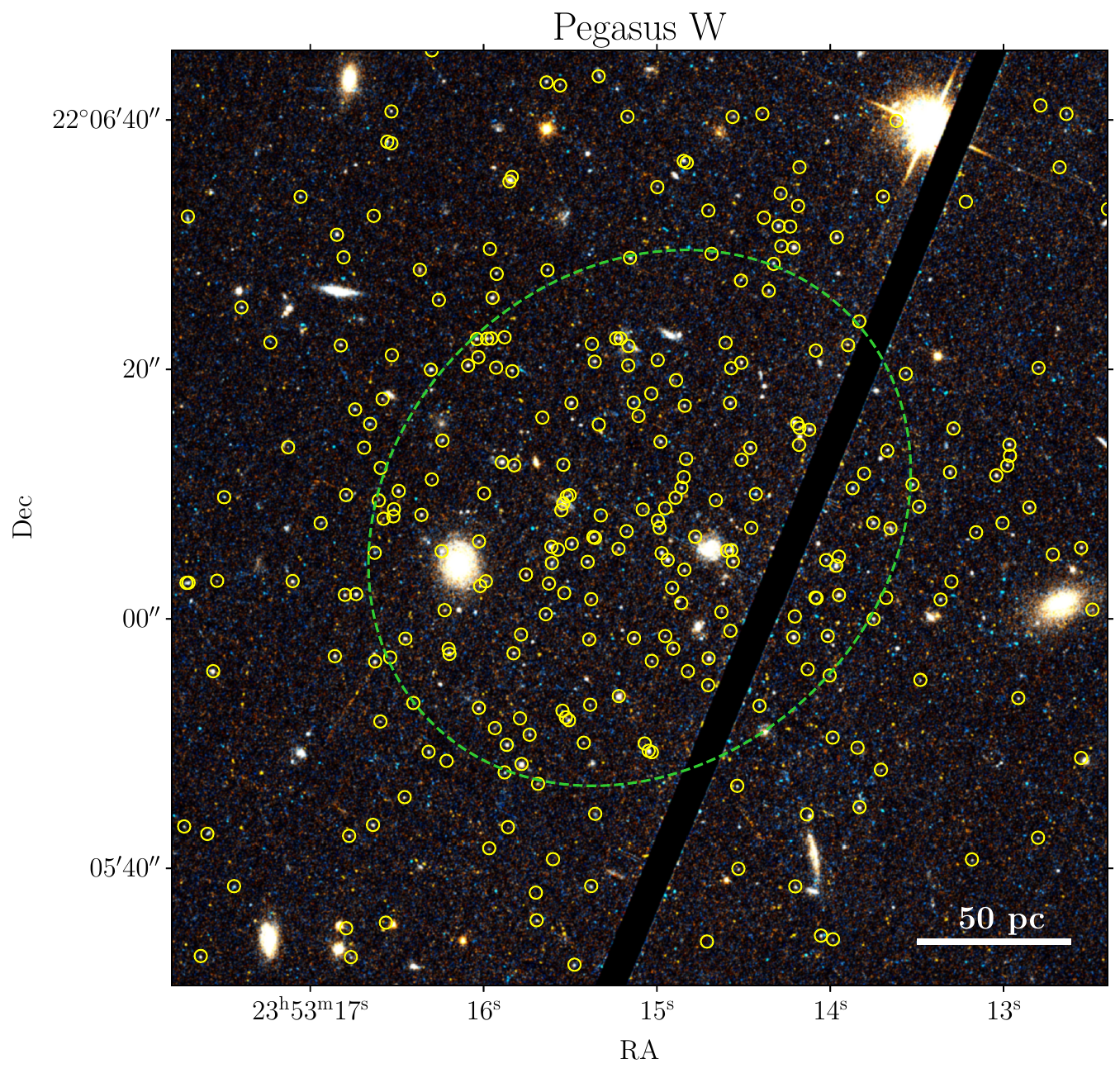}\\
    \includegraphics[width=0.49\linewidth]{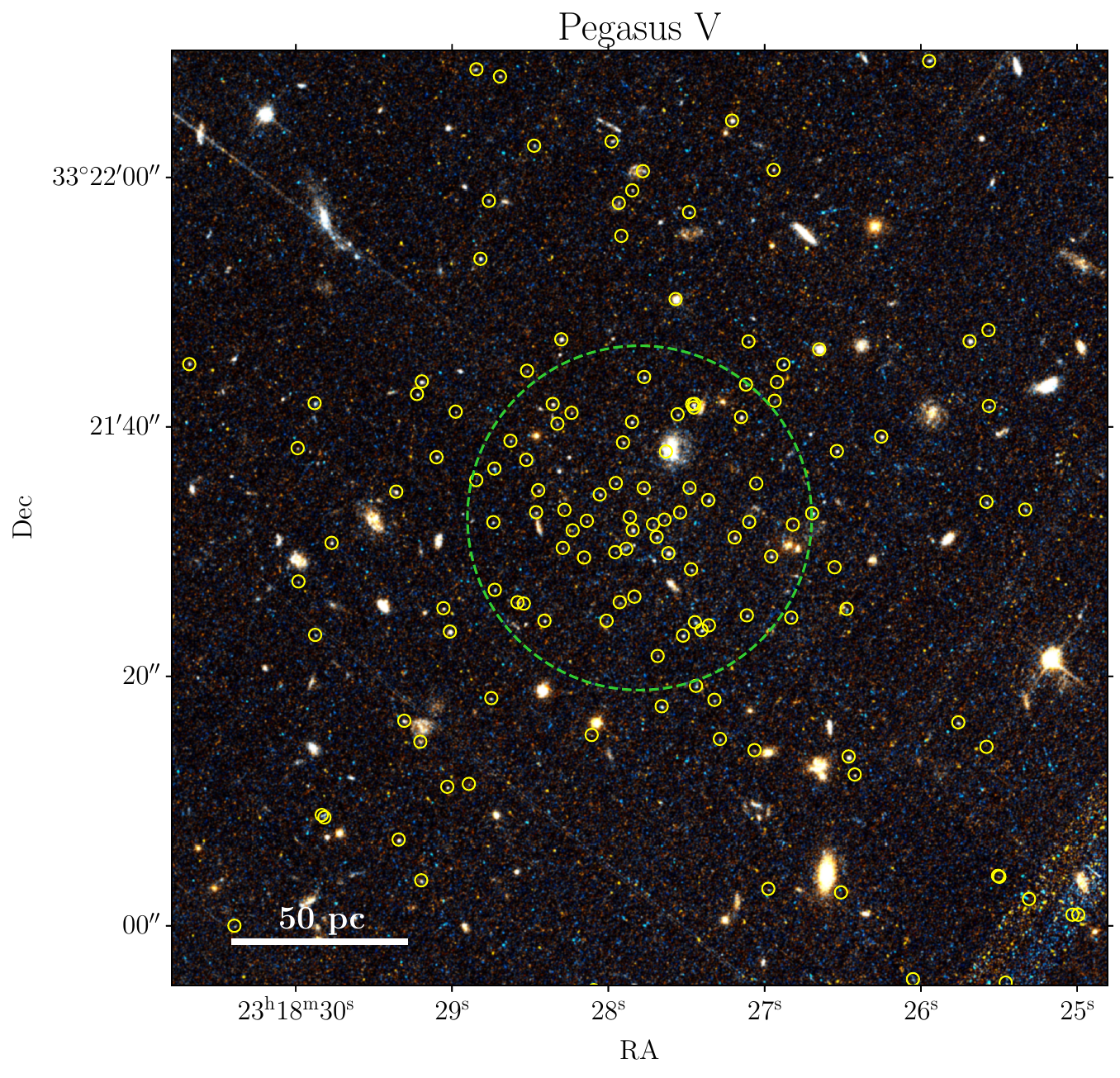}
    \includegraphics[width=0.49\linewidth]{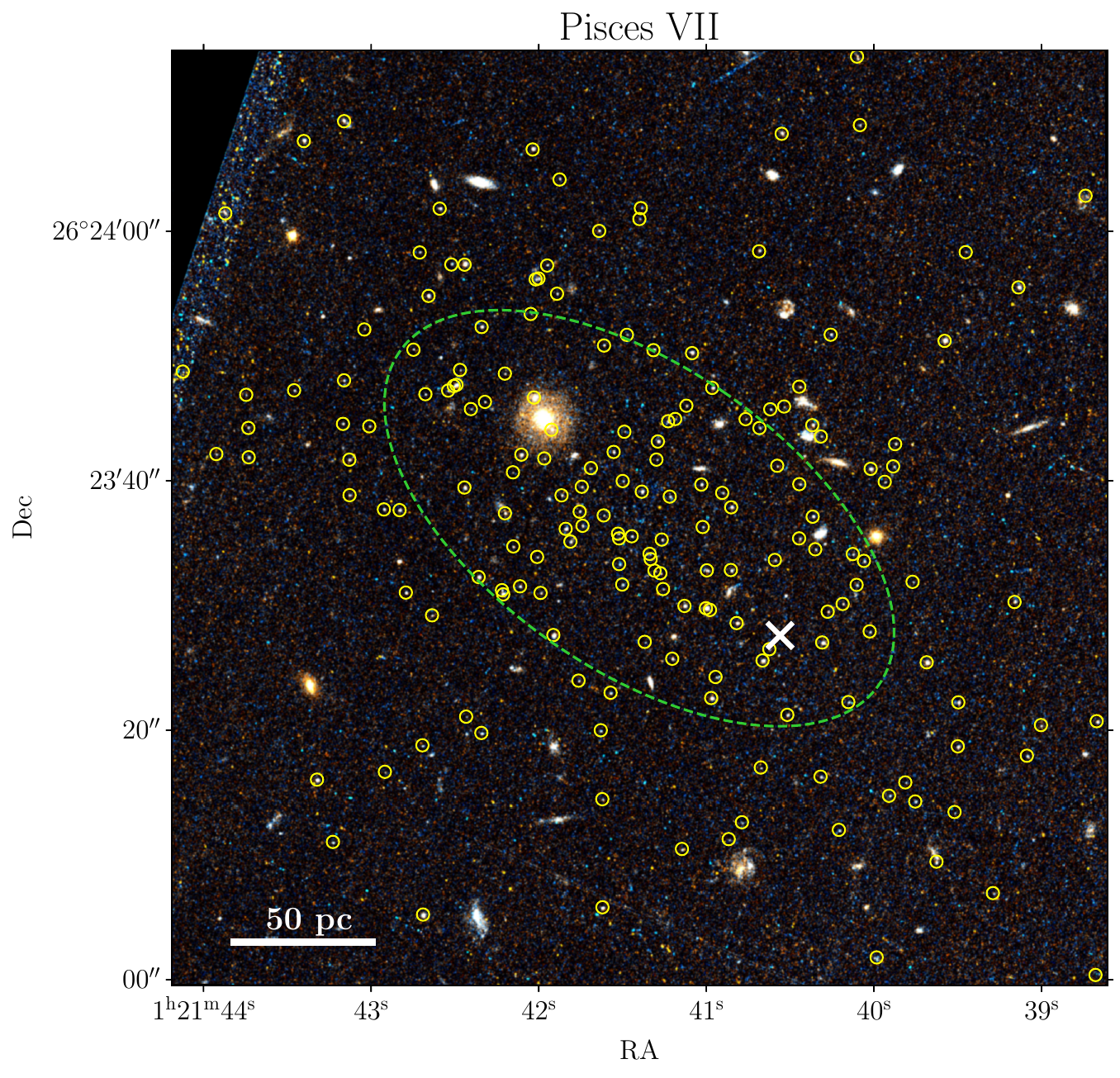}
    \caption{HST ACS F606W+F814W images of Pegasus~W (top), Pegasus~V (bottom-left) and Pisces~VII (bottom-right). Point sources that meet our RGB+HB selection criteria (\S\ref{sec:HSTdata}) are marked with yellow circles. The large dashed green ellipses indicate double the size of the half-light ellipse of each object. The large white X in the bottom-right panel indicates the center of Pisces~VII given in \citet{Collins+2024}. The background galaxy near the revised center likely adversely impacted the original ground-based photometry and led to an offset center being measured.}
    \label{fig:HST_cutouts}
\end{figure*}

HST Advanced Camera for Surveys/Wide Field Channel (ACS/WFC) F606W and F814W imaging of Pegasus~V and Pisces~VII was taken in January 2024 for the project GO-17316 (PI: M.~Jones). Each target was observed for a single orbit with exposures of approximately $\sim$1000~s in each filter. Very similar observations for Pegasus~W were taken previously (GO-16916; PI: K.~McQuinn), and we include these in this work in order to present a complete analysis of these three recently discovered UFDs in the direction of M31. Figure~\ref{fig:HST_cutouts} shows the false color composite of the F606W and F814W filters for all three UFDs. These specific data can be found at the Mikulski Archive for Space Telescopes via \dataset[10.17909/qfev-qm94]{https://doi.org/10.17909/qfev-qm94}.

All individual exposures for each target were processed with \texttt{DOLPHOT} \citep{Dolphin2000,dolphot} to produce a catalog of point sources. Quality cuts were made to eliminate background galaxies and spurious sources. We selected only type 1 and 2 sources (point and point-like sources) and sources with photometry flags equal to zero (higher flag values in \texttt{DOLPHOT} indicate sources overlapping chip edges, or containing many bad or saturated pixels). The \texttt{DOLPHOT} crowding, sharpness, and roundness parameters were restricted to maximum values of 1.0, 0.075, and 1.0, respectively. The completeness and photometric uncertainties for the resulting source catalog were estimated based on $10^6$ artificial stars, injected one at a time in random locations across the image and extracted with \texttt{DOLPHOT} (discussed further in Appendix~\ref{sec:SFHmethods}). Note that all quoted F606W and F814W magnitudes are in the Vega magnitude system and corrected for Galactic extinction following \citet{Schlegel1998} and \citet{Schlafly2011}. These corrections were implemented on a star-by-star basis using the \texttt{dustmaps} package \citep{Green+2018}. The median extinction values are given in Table~\ref{tab:props}.

For both targets member stars were initially selected based on previously published distances and structural parameters \citep{Collins+2022,Collins+2024}. The distance moduli were then adjusted manually until the red giant branch (RGB) in the color magnitude diagram (CMD) matched visually with a 13~Gyr old stellar population isochrone from the PAdova and TRieste Stellar Evolution Code \cite[PARSEC;][]{Bressan+2012}. RGB and horizontal branch (HB) stars were then selected around the PARSEC (RGB and HB) isochrone. All sources brighter than the 50\% completeness limit and with colors within the larger of $\pm0.12$~mag and $\pm$2$\sigma$ (where here $\sigma$ refers to the photometric uncertainties based on the artificial star tests) of the isochrone were selected. The minimum color width of the selection region ($\pm0.12$~mag) was chosen based on the width of the RGB in the CMD, so that the selection would not become unphysically narrow where the photometric uncertainties are small. The selected sources are highlighted with yellow circles in Figure~\ref{fig:HST_cutouts}, and are those used to determine our revised structural parameters in \S\ref{sec:struct_params}.

\begin{table*}
\centering
\caption{Properties of Pegasus W \& V, and Pisces~VII}
\begin{tabular}{cccc}
\hline\hline
                                          & Pegasus W & Pegasus V & Pisces VII \\ \hline
RA                                        &      23:53:15.1$\pm$2.2\arcsec     &   23:18:27.8$\pm$1.9\arcsec         &  01:21:41.4$\pm$2.4\arcsec          \\
Dec                                       &      +22:06:08.1$\pm$1.8\arcsec     & +33:21:32.7$\pm$1.9\arcsec          &  +26:23:37.0$\pm$2.1\arcsec          \\
$(m-M)$/mag                                 & $24.71\pm0.10$ & $24.39\pm0.15$ & $24.84\pm0.11$ \\
Dist./kpc                                 & $875\pm42$ & $755\pm52$ & $929\pm48$ \\
$D_\mathrm{M31}$/kpc   & 327 & 247 & 292 \\
$D_\mathrm{M33}$/kpc   & 361 & 413 & 105 \\
$\theta_\mathrm{M31}$  & 21.8$^\circ$ & 18.5$^\circ$ & 16.9$^\circ$ \\
$\theta_\mathrm{M33}$  & 24.0$^\circ$ & 28.7$^\circ$ & 5.0$^\circ$ \\
$m_\mathrm{F606W}$/mag & $17.70\pm0.14$ & $18.93\pm0.43$ & $18.59\pm0.17$ \\
$m_\mathrm{F814W}$/mag & $17.09\pm0.14$ & $18.30\pm0.38$ & $17.95\pm0.18$ \\
$M_V$/mag              & $-6.8\pm0.1$ & $-5.3\pm0.4$ & $-6.1\pm0.2$ \\
E(B-V)/mag & 0.11 & 0.07 & 0.11 \\
$r_h$/arcmin                                     &   0.39$\pm$0.03       &   0.23$\pm$0.05         &      0.39$\pm$0.05      \\
$r_h$/pc                                  & $99\pm9$ & $51\pm12$ & $105\pm15$ \\
$\epsilon$                                &     0.16$\pm$0.07     &  $<$0.22$^a$          &     0.48$\pm$0.07       \\
$a/b$                                     & $1.19\pm0.10$ & $\sim$1 & $1.92\pm0.26$ \\
$\theta$/deg                                 &   133$\pm$21        &   N/A        &     55.2$\pm$6.1       \\
$\log M_\ast$/\Msol        & $4.83^{+0.14}_{-0.06}$ & $4.18^{+0.16}_{-0.11}$ & $4.57^{+0.15}_{-0.07}$ \\
$\log M_\mathrm{HI}$/\Msol & $<$3.9       & $<$3.8       & $<$3.9        \\
$\log M_\mathrm{HI}/M_\ast$ & $<$$-$0.9       & $<$$-$0.4       & $<$$-$0.7        \\ \hline
\end{tabular}
\tablecomments{Separations from M31 and M33 are based on distances to these objects of 776.2 and 859.0~kpc, respectively \citep{Savino+2022}.\\$^a$Ellipticity limit is the 95\% confidence limit.}
\label{tab:props}
\end{table*}

\subsection{VLA \hi \ observations} \label{sec:VLAdata}

All three dwarfs were observed with the VLA in D-configuration during November 2023 for the project 23B-189 (PI: M.~Jones). Each target had an on source integration time of $\sim$2.8~h, split equally over two scheduling blocks. As the radial velocities of all three targets are unknown, the observations all used equivalent spectral setups with an 8~MHz ($\sim$1700~\kms) bandwidth centered at a frequency corresponding to a fiducial radial velocity of 200~\kms \ for the \hi \ line. This covers all plausible recessional velocities for these targets ($-600 < v_\mathrm{helio}/\mathrm{km\,s^{-1}} <  1000$). In each case the native channel width was 7.8~kHz ($\sim$1.7~\kms). 

These data were reduced using the VLA \hi \ pipeline of \citet{Jones+2023}, which makes use of standard Common Astronomy Software Applications \citep[\texttt{CASA};][]{CASA} tasks. Radio frequency interference (RFI) was excised using a combination of automated \texttt{CASA} tasks and manual flagging. The data were only mildly impacted by RFI, with about 10-20\% typically being flagged. However, one of the scheduling blocks for both Pegasus~W and Pegasus~V were more heavily affected, having $\sim$40\% of the data contaminated by RFI. Continuum subtraction was performed with a first order polynomial fit using line-free channels and during imaging the data were smoothed to a spectral resolution of 5~\kms. Imaging was performed using a Briggs robust parameter of 2.0 (i.e. roughly natural weighting) to optimize sensitivity to faint sources of comparable size to the synthesized beam. No CLEANing was performed as no significant sources of emission (other than the Milky Way) could be identified in the dirty image cubes.

\begin{figure*}
    \centering
    \includegraphics[width=0.7\linewidth]{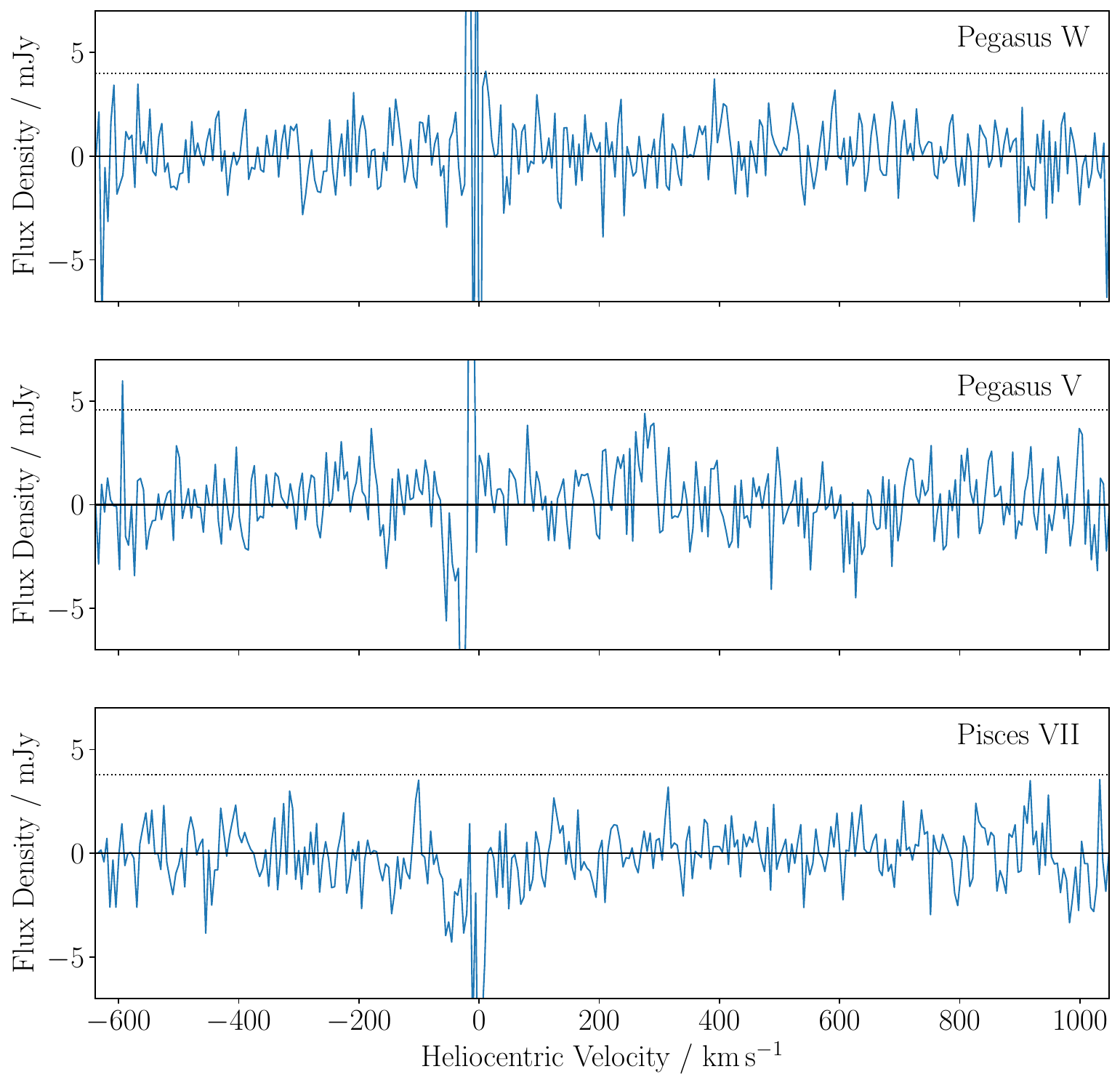}
    \caption{VLA \hi \ spectra of Pegasus~W (top), Pegasus~V (middle), and Pisces~VII (bottom) extracted over a single synthesized beam centered on the optical location of each object. The dotted horizontal line in each panel indicates three times the rms noise in the spectrum. None of the spectra contain significant signs of genuine line emission, aside from that of the MW around $v\approx0$~\kms. The low-significance feature in the Pegasus~V spectrum around 275~\kms \ is the result of large scale correlated noise features in the data.}
    \label{fig:VLAspecs}
\end{figure*}

Spectra were extracted within one synthesized beam centered on the optical position of each target (Figure~\ref{fig:VLAspecs}). The rms noise in each spectrum was calculated using the full bandwidth, excluding the band edges and $|v| < 100$~\kms. The rms values and beam dimensions are shown in Table~\ref{tab:VLAprops}. 

\begin{table}
\centering
\caption{VLA \hi \ observations}
\begin{tabular}{lccc}
\hline\hline
                      & Pegasus W & Pegasus V & Pisces VII \\ \hline
Beam                  & 65.8\arcsec$\times$54.9\arcsec \ & 65.6\arcsec$\times$51.9\arcsec & 71.4\arcsec$\times$55.4\arcsec \\
Position Angle        & 61.2$^\circ$ & 66.9$^\circ$ & 81.4$^\circ$ \\
$\sigma_\mathrm{rms}$/mJy/beam$^\dagger$ & 1.3 & 1.5 & 1.3 \\ \hline
\end{tabular}
\tablecomments{$\dagger$ The rms noise within 5~\kms \ channels.}
\label{tab:VLAprops}
\end{table}

\section{Distance and structural parameters} \label{sec:struct_params}

To measure the structural parameters of the three dwarf galaxies, we fit an exponential profile (on top of a background level) to the two-dimensional distribution of stars consistent with the RGB in each system using a maximum likelihood technique \citep{Martin08}, as implemented in \citet{Sand09,Sand12}.  Stars that are consistent with the RGB were chosen throughout the HST/ACS field of view for this analysis, down to F814W$\approx$26.5~mag (roughly corresponding to the 50\% completeness limit at $\mathrm{F606W-F814W = 0.5}$). The algorithm naturally accounts for regions of the image where point sources would not be detectable (such as near saturated stars).  The free parameters for the exponential fit are the central position ($\alpha_0$,$\delta_0$), position angle ($\theta$), ellipticity ($\epsilon$=$1-b/a$), half-light radius ($r_h$) and a constant background surface density.  Uncertainties on each parameter are derived via a bootstrap resampling analysis, run over 1000 iterations.  As an extra check on the results, the maximum likelihood analysis was re-run on the RGB star list with a magnitude cut 0.5 mag brighter than the full catalog; the structural parameters derived from this smaller RGB star list were consistent to within the uncertainties of the full analysis in all cases. The results of the structural analysis are shown in Table~\ref{tab:props}.

The values we derived for the structure of each dwarf are similar to those derived for these systems elsewhere in the literature.  First, we derive a nearly identical half-light radius and ellipticity for Pegasus~W as the discovery analysis did, using the same HST data set \citep{McQuinn+2023}, although we measure a $\sim$1$\sigma$ difference in the position angle of the dwarf.  For Pegasus~V and Pisces~VII, our HST data are significantly higher resolution and deeper than the ground-based discovery data.  For Pisces~VII, we have significantly revised the central position by $\approx$15\arcsec \ (Figure~\ref{fig:HST_cutouts}, bottom-right) and half-light radius of the dwarf (smaller by $\approx$60\%).  We attribute this purely to the improved fidelity of our HST-based data set, as Pisces~VII was not well characterized in the discovery data.

\begin{figure*}
    \centering
    \includegraphics[width=\textwidth]{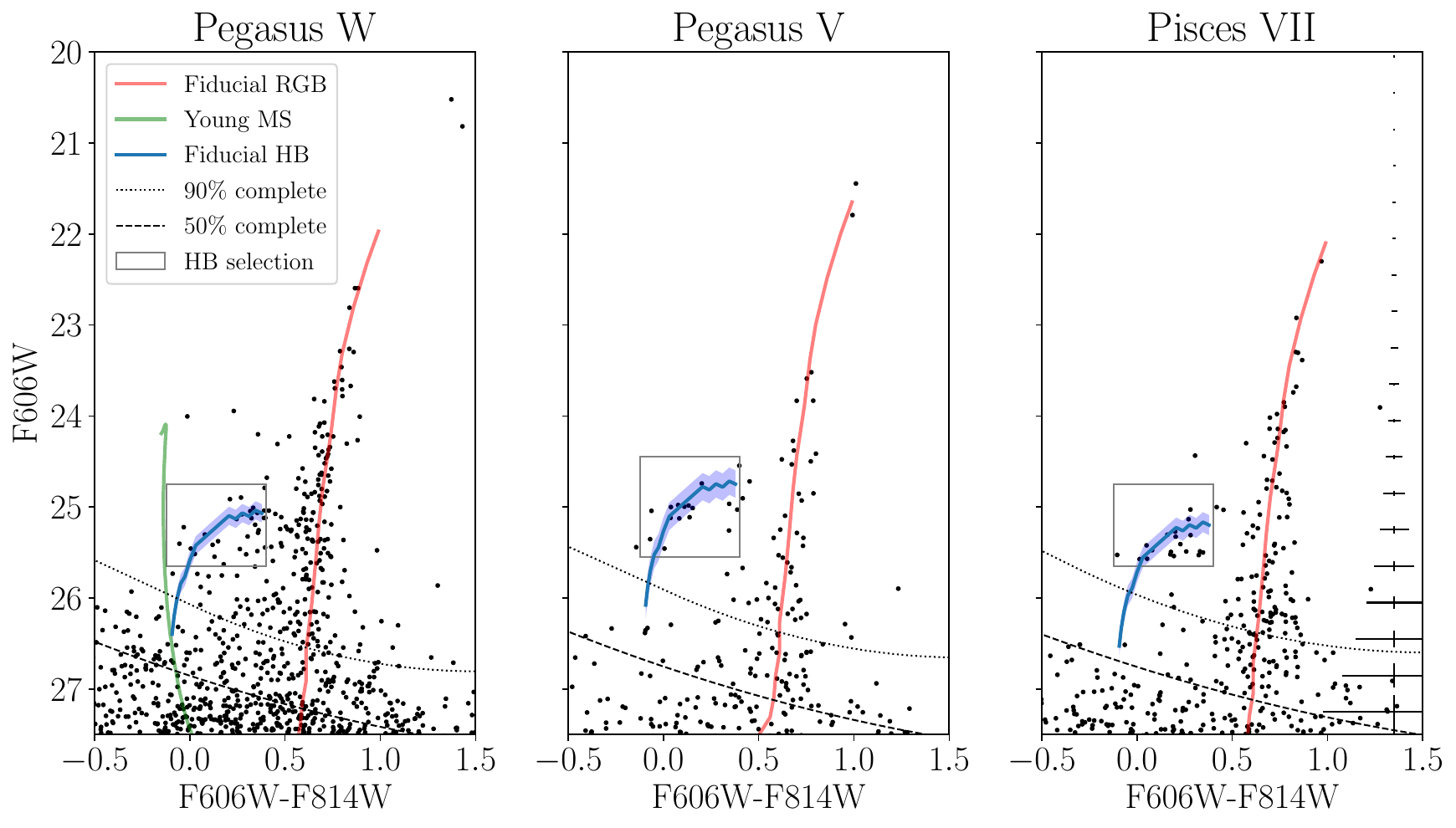}
    \caption{HST CMDs (extinction-corrected) of Pegasus~W (left), Pegasus~V (center), and Pisces~VII (right) within $2r_h$. The blue and red lines show the fiducial HB and RGB \citep[from M92;][]{Mutlu-Pakdil+2019}, respectively, while the green line shows a young MS population from PARSEC ($\log t/\mathrm{yr} = 8.3$, $[\mathrm{M/H}] = -1.7$), all shifted to the appropriate distance moduli (Table~\ref{tab:props}). Almost no stars are present at the TRGB and thus we instead determined distances from the location of the HB (\S\ref{sec:struct_params}). The stars used to fit the distance moduli are those within the HB selection boxes (grey rectangles). The blue shaded bands indicate the 1-$\sigma$ uncertainties in the distance moduli. The dotted and dashed black lines indicate the 90\% and 50\% completeness limits, respectively, for each observation. The errorbars along the right edge of the right panel indicate the typical photometric uncertainties as a function of F606W magnitude, which are comparable for all three observations.}
    \label{fig:CMDs}
\end{figure*}

Final CMDs for each object show the stars within $2r_h$ (Figure~\ref{fig:CMDs}). Due to the sparsely populated RGBs of all three targets, using the tip of the RGB as a standard candle for distance determination was not viable \citep[cf.][]{Mutlu-Pakdil+2022}. Instead we use the clear HBs that are visible in all three CMDs.

We first experimented with using a fixed HB magnitude calibration \citep[e.g.][]{Carretta+2000}, as \citet{McQuinn+2023} did for Pegasus~W. However, neither Pegasus~V nor Pisces~VII have many redder HB stars where the HB is flattest and such calibrations are most applicable. We therefore adopted the M92 HB \citep[determined with HST imaging by][]{Mutlu-Pakdil+2019} as a fiducial to compare with our CMDs. For each target we created a rectangular HB selection box in the CMDs (grey rectangles in Figure~\ref{fig:CMDs}). The color extent of the selection was set by the range covered by the M92 fiducial HB, while the F606W magnitude extent was determined visually. For both Pegasus~V and Pisces~VII the HB is quite distinct from the rest of the population, and the appropriate vertical extent was clear. However, for Pegasus~W it is less clear as there is a nearly continuous population of stars from the completeness limit to the HB (likely an indication of a more complex SFH). Using the 2-dimensional photometric uncertainty model (described in \S\ref{sec:phot_uncert}) and a maximum likelihood fitting approach, we determined the best fit distance modulus for each target by maximizing the agreement between the observed HB population and the fiducial curve. Following \citet{Mutlu-Pakdil+2019}, the distance modulus of M92 is taken as the average of the measurements in \citet{Paust+2007}, \citet{Brown+2014}, and \citet{Sollima+2006}, giving $(m-M)_\mathrm{M92} = 14.62\pm0.06$. To estimate the uncertainties we combined in quadrature bootstrap uncertainties for the best fit distance modulus, the differences in the fits for stars within 1$r_h$ and 2$r_h$, and the uncertainty in $(m-M)_\mathrm{M92}$. This likely still slightly underestimates the true uncertainty as aspects such as the difference in metallicity compared to M92 and how the HB is populated are not taken into account. 

Our final distance estimates and their uncertainties are given in Table~\ref{tab:props}. The distance we measure for Pegasus~W is a few percent smaller than that determined by \citet{McQuinn+2023} using the same data (but a different approach), however, these values still agree within their uncertainties. For Pegasus~V and Pisces~VII previous distance estimates were determined from ground-based data and were therefore more likely to suffer from contamination and larger photometric uncertainties. Our distance for Pegasus~V is $\sim$2$\sigma$ ($\sim$70~kpc) larger than that given in \citet{Collins+2022}, while the distance for Pisces~VII agrees within the uncertainties \citep{Collins+2024}.

\section{Star formation history fitting} \label{sec:SFHfitting}

CMD-based SFH fitting codes \citep[e.g.,][]{Harris+2001,MATCH,IAC-pop,Garling+2025,PANCAKE} attempt to extract information about past episodes of SF by fitting linear combinations of single stellar populations (SSPs) to the populations of stars remaining in the present day CMD of a resolved galaxy or star cluster. Although these SSPs are not orthogonal to each other and the full parameter space has a potentially enormous dimensionality, with careful constraints on, and binning of, the age and metallicity of the SSPs, it is possible to reduce this to a relatively straightforward maximum likelihood problem with a global maximum that is readily identifiable with standard optimization algorithms. 

Our fitting methodology is modeled on that of \texttt{StarFISH} \citep{Harris+2001,StarFISH}, which was originally developed in \texttt{FORTRAN}. We have mostly followed the steps of \texttt{StarFISH}, but these are now implemented in \texttt{Python}. In this section we provide a brief description of the key points of this fitting process and direct readers interested in the details of our approach to Appendix~\ref{sec:SFHmethods}. In Appendix~\ref{sec:mockSFHs} we perform various mock fits and give an indication of the reliability of our SFHs given the photometric depth of the HST data. 

Based on three isochrone sets, PARSEC \citep{Bressan+2012}, MESA Isochrones \& Stellar Tracks \citep[MIST;][]{Dotter+2016,Choi+2016}, and the Bag of Stellar Tracks and Isochrones \citep[BaSTI;][]{Pietrinferni+2004,Hidalgo+2018}, we construct SSP binned CMDs (or ``Hess diagrams") over a grid of stellar ages and metallicities. These incorporate realistic photometric uncertainties and completeness based on artificial star tests conducted separately for each observation. These SSP Hess diagrams span the color range $-0.5 < \mathrm{F606W-F814W} < 1.5$ and the magnitude range $18 < \mathrm{F814W} < 27$, with the faint limit roughly corresponding to the 50\% completeness limits of our observations (although stars in the SSPs are generated down to 1~mag fainter than this to ensure there are no edge effects). 

The real CMDs are then binned and fit with linear combinations of the SSPs (with each isochrone set fit independently) in order to determine the SFHs. The random uncertainties in the fits are estimated by performing Monte Carlo re-sampling of the best-fit CMDs and then repeating the fitting process. However, the systematic uncertainties resulting from differences in the stellar evolution models are likely the dominant source of uncertainty. We assess the scale of these systematic uncertainties by fitting with three distinct stellar libraries (Appendix~\ref{sec:SFHmethods} \& \ref{sec:mockSFHs}). \citet{Dolphin2012} argues that, although this approach is widely used, it can lead to the underestimation of systematic uncertainties, and we plan to explore more comprehensive approaches in future work.

\section{Results} \label{sec:results}

\subsection{Star formation histories}

The cumulative SFHs resulting from the fitting process described in Appendix~\ref{sec:SFHmethods} are shown in Figure~\ref{fig:cumSFHs} and normalized by the mass of all stars ever formed in each galaxy. Figure~\ref{fig:diffSFHs} shows the differential SFHs of the same fits with a logarithmic time axis.

\begin{figure*}
    \centering
    \includegraphics[width=0.9\textwidth]{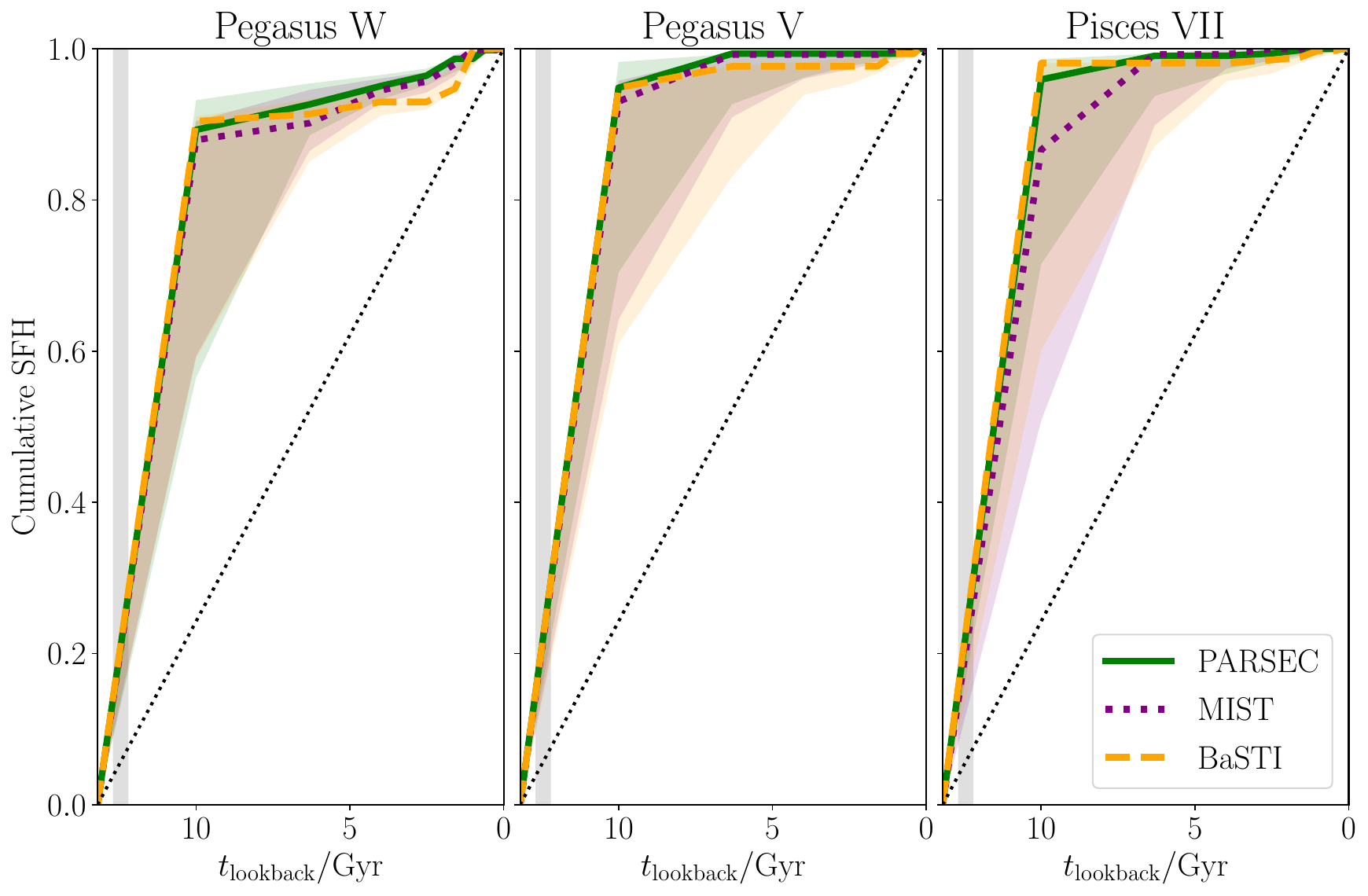}
    \caption{Cumulative fractional star formation histories of Pegasus~W (left), Pegasus~V (center), and Pisces~VII (right). The fits from the three different isochrone libraries are shown with different colors and line styles. The diagonal dashed black lines represent a constant SFR for the age of the universe and the grey vertical band indicates the approximate period of cosmic reionization ($10 \lesssim z \lesssim 6$). The shaded color bands indicate the 1-$\sigma$ uncertainties of each model based on re-fitting MC realizations of the best fit artificial CMD. Note that the first time bin extends from the beginning of the universe until 10~Gyr ago. As this is a single bin, the SFR within it is plotted at the average (constant) value. In reality, the majority of the stars formed in this bin likely formed by the end of reionization (grey band). However, the current photometry is not deep enough to distinguish these ancient populations and they are thus grouped into a single bin.}
    \label{fig:cumSFHs}
\end{figure*}

\begin{figure*}
    \centering
    \includegraphics[width=\textwidth]{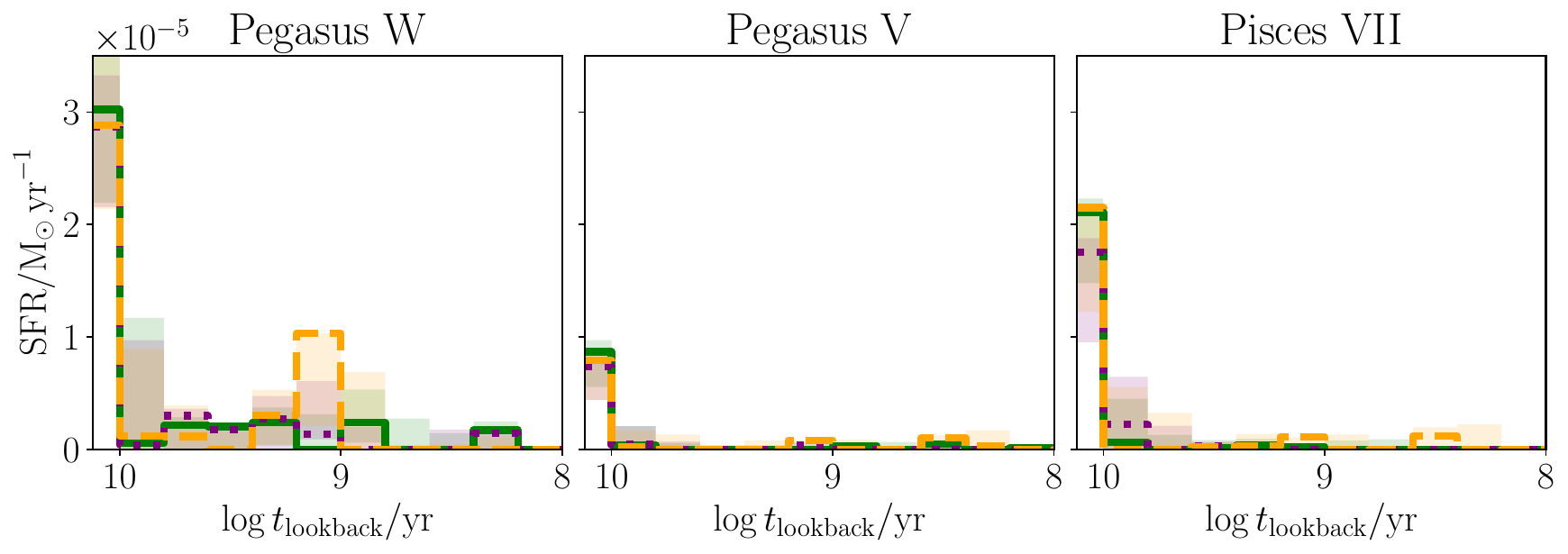}
    \caption{Differential star formation histories of the three UFDs, plotted in a style analogous to Figure~\ref{fig:cumSFHs}. Only Pegasus~W exhibits significant SF past 10~Gyr ago. We have truncated the SFHs at 100~Myr as SFRs more recent than this are unreliable given the sparsely populated CMDs of these UFDs. 
    Note that the error bars on differential SFHs can be inflated by anti-correlations between adjacent time bins, causing them to overestimate the true statistical uncertainties. Such correlations do not significantly impact error estimates for the cumulative SFHs, thus the error bands in Figure~\ref{fig:cumSFHs} provide a more robust representation of the uncertainties. }
    \label{fig:diffSFHs}
\end{figure*}

Our SFH of Pegasus~W is qualitatively similar to that of \citet{McQuinn+2023}. In particular, we reproduce the striking finding that the best fit SFH exhibits sustained SF in all three models until $\sim$1~Gyr ago, and perhaps even more recently. Meanwhile both Pegasus~V and Pisces~VII show SFHs more typical of other UFDs, with the vast majority of their stars formed in the first time bin of the SFH. Although this first bin is very broad (a result of the depth of the current photometry), this is consistent with what would be expected for objects that are quenched as a result of cosmic reionization. There are some later time blips of SF evident in the differential SFHs (Figure~\ref{fig:diffSFHs}) of these two UFDs, but such features result in so few stars in the present day CMD that they should be treated with considerable caution. Likely they are the result of small groups of stars that the template CMDs struggle to fit (such as blue stragglers or uncommon contaminants stars that are not  adequately represented in the contaminant CMD), rather than significant late-time SF. This will be discussed further in \S\ref{sec:PegW_SFH}.

\citet{Collins+2024} suggested that Pisces~VII might contain young Helium-burning (HeB) stars because several sources in their ground-based CMD are slightly brighter than the HB (similar to Pegasus~W's CMD; Figure~\ref{fig:CMDs}, left). However, our CMD of Pisces~VII (Figure~\ref{fig:CMDs}, right) shows only one such star and suggests that those identified previously were likely the result of contaminants that the improved depth and resolution of HST imaging have removed. The CMD of Pegasus~V shows none of these stars. By contrast, both Pisces~VII and Pegasus~V show a small number of potential blue stragglers blueward and below the HB, and this is likely responsible for the minimal SF in the past several Gyr of their SFHs (Figure~\ref{fig:diffSFHs}), as these can be erroneously modeled with young stars in the fitting process. The implications of these stars will be discussed in detail for Pegasus~W in \S\ref{sec:PegW_SFH}, which has by far the largest population of the three UFDs.

Although it is common to estimate quenching times (e.g. $\tau_{90}$ or $\tau_{80}$) marking when galaxies have formed a certain fraction of their stars (e.g. 90\% or 80\%), in most cases for our best fit SFHs such a threshold would be surpassed in the first bin and we have therefore elected not to attempt to quantify quenching times. 
Both Pegasus~V and Pisces~VII appear to have formed $\gtrsim$90\% of their stars by 10~Gyr ago. Whereas Pegasus~W appears to have formed just shy of 90\% by that stage, after which it continued to form stars at a much reduced (but roughly constant) rate, until less than 1~Gyr ago. This is qualitatively similar to the SFH history calculated for Pegasus~W by \citet{McQuinn+2023}, based on the same observations.
Deeper observations that better constrain the ancient populations of these galaxies would be needed to reliably determine quenching times. Instead in \S\ref{sec:dicussion} we will discuss how the SFHs of these three objects compare qualitatively to other UFDs in the M31 and MW systems.

\subsection{Stellar masses \& luminosities}

Determining present day stellar masses and luminosities from the SFHs is non-trivial and is discussed in detail in Appendix~\ref{sec:SFH_masses}. We follow a similar approach to \citet{McQuinn+2024} for estimating stellar masses, assuming a recycling factor of 43\%, while to estimate luminosities we develop our own approach that sums up the contribution of every SSP in the SFH.

Our stellar mass estimates for the three UFDs can be found in Table~\ref{tab:props}. The estimate for Pegasus~W ($\log M_\ast/\mathrm{M_\odot} = 4.83$) is almost identical to that of \citet{McQuinn+2023}. This near-perfect agreement is made slightly worse after accounting for the fact that we adopted a different distance than \citet{McQuinn+2023}, but even so, the disagreement is still smaller than the quoted uncertainties in either measurement. This stellar mass estimate places Pegasus~W near the boundary thought to separate UFDs from galaxies that can survive cosmic reionization and continue forming stars. Our stellar mass estimates for both Pegasus~V and Pisces~VII are considerably lower, placing them comfortably within the mass range that would be expected to be quenched by reionization.

The luminosities in F606W and F814W are calculated directly from the best fit SFHs (see Appendix~\ref{sec:SFH_masses}) and we then convert these to apparent magnitudes ($m_\mathrm{F606W}$ \& $m_\mathrm{F814W}$; Table~\ref{tab:props}) based on our distance moduli measurements. These magnitude therefore intrinsically include corrections for stars below the completeness limit.
We convert these to V-band following \citet{Sirianni+2005}, and all three values are quoted in Table~\ref{tab:props}. Our $M_V$ estimate for Pisces~VII is consistent with that measured by \citet{Collins+2024}. However, the value for Pegasus~V differs from \citet{Collins+2022} by a full magnitude. It is unclear what is causing this discrepancy, but we suspect it may be a result of considerably worse field contamination in the ground-based imaging used by \citet{Collins+2022}, compared to the HST imaging used here. We also find a slightly different magnitude for Pegasus~W \citep[0.3~mag fainter than][]{McQuinn+2023}. This is likely a systematic difference resulting from the two approaches taken to estimating the luminosity.

\subsection{\hi \ mass limits}

No significant signs of line emission were identified in the spectra shown in Figure~\ref{fig:VLAspecs}, nor was any evident when the \hi \ data cubes were visually inspected. Although there are some $\sim$2$\sigma$-level peaks in these spectra (in particular, at $\sim$300~\kms \ for Pegasus~V and -100~\kms \ for Pisces~VII), visual inspection of these faint features strongly suggest that they are due to structured noise in the cube, not real emission features.

We estimate the maximum flux that could be present in a top hat-shaped line profile with an assumed width of 25~\kms, similar to that of Leo~P \citep{Giovanelli+2013}, and not exceeding the 3$\sigma$ level plotted in Figure~\ref{fig:VLAspecs}. These 3$\sigma$ upper limits were converted to \hi \ mass upper limits based on the distance to each UFD using the standard conversion \citep[e.g.][]{Giovanelli+2005}. These limits are given in Table~\ref{tab:props}.
All of the UFDs have stringent upper limits on their \hi \ content ($M_\mathrm{HI} < 10^4$~\Msol) or, equivalently, their gas fractions ($M_\mathrm{HI}/M_\ast$), all of which are constrained to be well below unity. Thus, these are all gas-poor objects. The only exception might be if their (unknown) radial velocities are near zero as then their \hi \ content could exceed the above limit, but be overwhelmed by the \hi \ emission of the MW.

\section{Discussion} \label{sec:dicussion}

The typical quenching times of MW and M31 satellites are known to differ, with many MW satellites quenching early on (and the remainder either still being star-forming, or quenching in the past few Gyr), while many M31 satellites have more extended SFHs \citep[e.g.][]{Skillman+2017,Weisz+2019}. This has been interpreted \citep[e.g.][]{Weisz+2019,D'Souza+2021} as evidence that the satellite system (and stellar halo) of M31 was built up in a single epoch centered around 6 Gyr ago, resulting in quenching of new satellites that had previously been in the field. Meanwhile they suggest that the MW stellar halo and satellite system was built up in two epochs, one $\sim$10~Gyr ago, corresponding to the Gaia-Enceladus dwarf accretion event \citep{Helmi+2018,Chaplin+2020}, and another $\sim$1~Gyr ago when the LMC fell into the LG accompanied by its own satellites \citep{Besla+2010,Kallivayalil+2018,Patel+2020}. Thus, an early quenching epoch is expected for most MW satellites (as well as some very recent quenching), while most M31 satellites are expected to have quenched later.

This argument is somewhat circular as \citet{D'Souza+2021} determined the times of accretion episodes based on the quenching times of satellites. However, only the higher mass satellites were used to do this, and it is therefore worthwhile to investigate whether lower mass satellites, such as UFDs, follow this same trend or if they were quenched earlier by reionization. In the MW system efforts are confounded by the fact that the first epoch of quenching ($\sim$10~Gyr ago) is not long after the end of reionization, especially given the temporal resolution of the typical SFHs. However, in the M31 system UFDs genuinely quenched by reionization should stand out from the majority of other satellites that were quenched by environmental processes after joining the M31 system.

\subsection{Comparison to other M31 UFDs}

\begin{figure}
    \centering
    \includegraphics[width=0.75\linewidth]{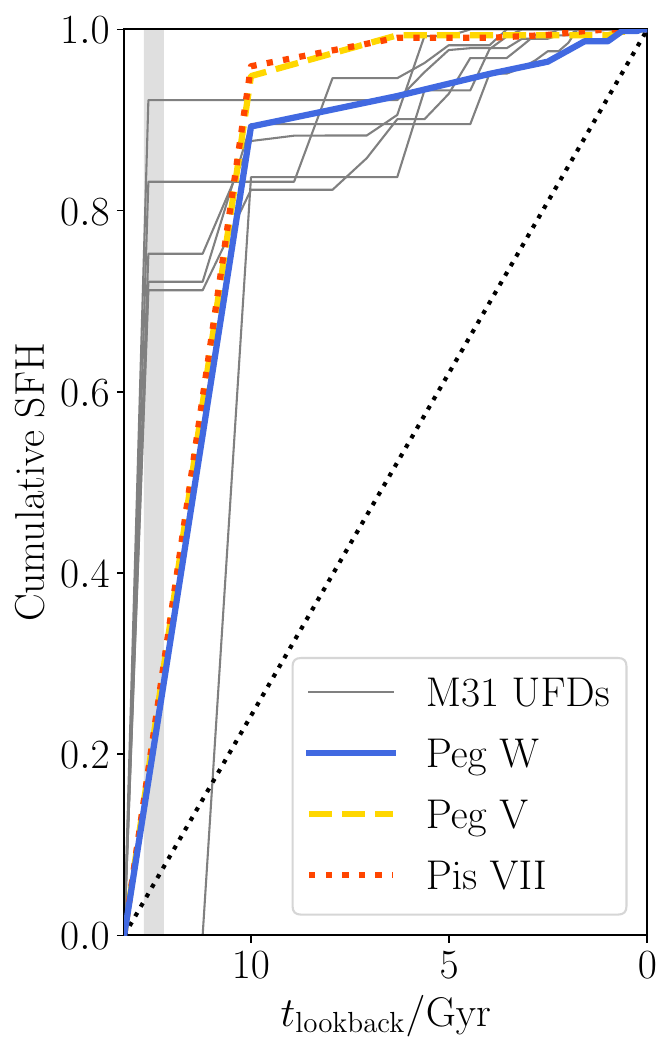}
    \caption{Cumulative fractional star formation histories (for PARSEC isochrones) of Pegasus~W (blue solid line), Pegasus~V (gold dashed line), and Pisces~VII (peach dotted line) in comparison to the six M31 UFDs (thin grey lines) from \citet{Savino+2023}. Other markings as in Figure~\ref{fig:cumSFHs}. Note that the first bin of our SFHs is considerably wider than those of \citet{Savino+2023} owing to our shallower photometry; apparent differences beyond $\sim$10~Gyr ago are therefore not meaningful.}
    \label{fig:cumSFH_M31_UFDs}
\end{figure}

\citet{Savino+2023} fit SFHs to six fairly bright ($-7 \lesssim M_V \lesssim -6$) UFDs near M31 using deep HST photometry reaching down to the oMSTO. They found that the UFDs were heavily impacted by reionization, with long pauses in their SFRs, but were not ultimately quenched by it. Instead all six appear to have formed a significant fraction of their stars long after reionization. This sample contained mostly objects that are comparable in luminosity to Pegasus~W and reaching as faint as Pisces~VII, but typically $>$1~mag brighter than Pegasus~V (see magnitudes in Table~\ref{tab:props}). 

In Figure~\ref{fig:cumSFH_M31_UFDs} we show the PARSEC-based SFHs of our three target UFDs in comparison to the six UFDs fit in \citet{Savino+2023}. The deeper photometry of \citet{Savino+2023} allows for finer temporal binning and more reliable SFHs at ancient times. Thus, the details of the SFHs beyond 10~Gyr ago are obscured for our targets \citep[relative to those of][]{Savino+2023}, because of the depth of the single-orbit HST observations and the temporal binning. However, our tests with mock populations (Appendix~\ref{sec:mockSFHs}) indicate that the SFH fits are generally reliable back to at least $\sim$7~Gyr ago.

Setting aside the differences at early times resulting from the reasons discussed above, our SFH of Pegasus~W appears quite consistent with these other M31 UFDs until the past few Gyrs. At this point all of the six comparison UFDs have entirely ceased forming stars, whereas Pegasus~W continues until $\sim$1~Gyr ago. Five of the six comparison UFDs are only $\sim$100~kpc away from M31 and have likely been satellites of M31 for some time \citep{Savino+2023,Savino+2025}. If Pegasus~W is on its first infall into the M31 system then it may be analogous to these six other systems, but environmental effects did not truncate its SF until the past Gyr. The detailed interpretation of Pegasus~W's SFH is discussed further in \S\ref{sec:PegW_SFH}.

In contrast to Pegasus~W and the other six M31 UFDs, Pegasus~V and Pisces~VII both form well over 90\% of their stellar mass within the first bin of our SFHs. As these SFHs cannot be considered reliable at the few percent level, this is consistent with them having quenched during or shortly after reionization. Their cumulative SFH curves fall consistently above those of the six comparison UFDs, and appear to be more akin to those of faint UFDs near the MW \citep{Brown+2014,Weisz+2014,Sacchi+2021,Durbin+2025}. Indeed, \citet{Savino+2025} suggest that the previously known UFDs around M31 might be too massive to have been fully quenched by reionization alone, and the lack of lower mass UFDs known around M31 may have contributed to the prolonged periods of SF observed (post-reionization) relative to the MW UFDs. Our tests with mock SFHs (Appendix~\ref{sec:mockSFHs}) indicate that, with the depth of the current photometry, our analysis cannot always reliably distinguish between purely ancient stellar populations and the SFHs of other M31 UFDs \citep{Savino+2023}, however, the rapid rise of the SFHs fitted to Pegasus~V and Pisces~VII places them in a fairly unambiguous part of the parameter space that is distinct from the \citet{Savino+2023} UFDs (see discussion in Appendix~\ref{sec:mockSFHs}). Thus, Pegasus~V and Pisces~VII appear to fill the mass gap for UFDs near M31, and support the assertion that at lower masses the M31 UFD population may well be more like that of the MW's, with galaxies quenching at early times, likely due to reionization \citep[e.g.][]{Brown+2012,Brown+2014,Sacchi+2021}.

\subsection{The extended SFH of Pegasus~W} \label{sec:PegW_SFH}

The prolonged SFH derived for Pegasus~W appears to be driven by two unusual populations of stars in its CMD: a small group about 1~mag brighter than the HB, and a significant cloud about 1--2~mag fainter than the HB, extending to very blue colors. 
In the modeled SFH the former appear to be fit with HeB and red clump (RC) stars (less than a few Gyr old). If we manually remove them from the input CMD then the PARSEC-based SFH does not form any stars in the range $9.0 < \log t_\mathrm{lookback}/\mathrm{yr} < 9.4$. Although there is some chance that these stars could be contaminants, they are fairly bright and there are no other equivalent (in color and magnitude) sources throughout the entire FoV, they all lie within 2$r_h$ of Pegasus~W's center.  

The other population, the excess of stars below the HB and blueward of the RGB, seem to be fit with main sequence (MS) stars from even younger ($<$1~Gyr) populations in our SFHs. 
Although there a more contaminants at these fainter magnitudes, these sources also appear to be concentrated at Pegasus~W's location, and are thus unlikely to be solely the result on contaminants.\footnote{Upon inspection of the faintest sources in the Pegasus~W imaging, we noted an unusually large number of faint spurious sources at $\mathrm{F606W} > 26.5$ that were not uniformly distributed across the FoV, but clustered towards the chip gap. Although these spurious sources were already accounted for in our template background CMD (Appendix~\ref{sec:binned_mock_CMDs}) and many are excluded from our SFH analysis (which only fits sources $\mathrm{F814W} < 27$), we experimented with re-fitting Pegasus~W's SFH with a limiting magnitude of $\mathrm{F814W} = 26.5$. This resulted in SFHs with the same general features seen in Figure~\ref{fig:cumSFHs} (left), including recent SF.} \citet{McQuinn+2023} noted that the recent SF implied by this population might be the result of the fitting process mistaking blue stragglers (which are not included in the isochrone sets) for MS stars. Blue stragglers, which are thought to be the result of collisions or mass transfer between old stars, can be approximately modeled with young metal-poor isochrones, as post-merger they are expected to behave similarly to young, low metallicity MS stars. Figure~\ref{fig:CMDs} (left) shows a young PARSEC isochrone (green line) approximately matching the age and metallicity of the last significant burst of SF modeled for Pegasus~W in the MIST and PARSEC SFHs.

Blue stragglers are a relatively common occurrence in the CMDs of other low-mass dwarfs and there is an observational trend between the ratio of the number of stars on the HB and the number of blue stragglers, and the V-band luminosity of a galaxy \citep[e.g.][]{Momany+2007,Sand+2010}. This relation \citep{Monelli+2012} suggests that for Pegasus~W there should be approximately 1.5 times as many blue stragglers as HB stars, which we find is roughly consistent with the number that we see (although there is considerable scatter in this relation). However, it is worthwhile to briefly consider the alternative, that is, that (some of) these stars are genuinely indicative of recent SF.

Taken at face value, the modeled SFH of Pegasus~W indicates that it was strongly impacted by an event $>$10~Gyr ago (likely cosmic reionization), but was able to resume accreting gas and continue SF, similar to some more massive dwarfs \citep{Cole+2007,Cole+2014,McQuinn+2024,McQuinn+2024b}. We note that Pegasus~W's stellar mass ($\log M_\ast/\mathrm{M_\odot} = 4.83$) is very close to the commonly assumed approximate threshold mass for UFDs, $M_\ast \approx 10^5$~\Msol \ \citep[e.g.][]{Simon+2019}. Perhaps Pegasus~W is in fact right at this threshold and although its potential well was insufficient to stave off the effects of reionization completely, it was able to either retain gas, or subsequently accrete it, and continue forming stars. However, to conclusively resolve the nature of Pegasus~W, deeper observations are needed. Both spectroscopic and deep imaging observations of Pegasus~W are scheduled for JWST Cycle 4 (JWST-GO-7196 and JWST-GO-7119). If the population of blue stars below the HB really is due to blue stragglers and scattered RGB stars (with large photometric uncertainties) in the HST images, then these JWST observations should simply show a blue plume of stars extending above the main sequence turn-off (MSTO) of a simple ancient population.

\subsection{Impact of Environment}

Recent models of cosmic reionization have considered the impact of patchy reionization on the quenching of UFDs \citep[e.g.][]{Simpson+2013,Kim+2023}, finding that local variations in the timing of reionization can have considerable knock-on effects for galaxy quenching times. Observational studies of MW (and LMC) UFD satellites \citep{Sacchi+2021,Durbin+2025} found significant differences in the quenching times of UFDs associated with the MW, the LMC, and those on first infall, and suggest that environment at the time of reionization could be the cause. However, these variations are on the order of 1~Gyr and do not appear to be capable of explaining the difference in the SFHs of Pegasus~V and Pisces~VII versus Pegasus~W and other M31 UFDs \citep{Savino+2023}. On the other hand, in their zoom-in cosmological simulations \citet{Christensen+2024} also study the impact of environment and reionization, finding, contrary to prior expectations, that $\sim$20\% of UFDs with no massive neighbor within a Mpc are star-forming at $z=0$. Although none of our targets are this isolated, they could have been in the recent past. For example, if Pegasus~W is on its first infall and approaching M31 at the relatively low velocity of 200~\kms \ (note that the radial velocity of Pegasus~W is unknown and this value is purely for the sake of discussion), then 3-4~Gyr ago it would have been a Mpc away. Thus, while Pegasus~V and Pisces~VII appear to have quenched at ancient times, there is a theoretical basis for the hypothesis that Pegasus~W might have managed to maintain SF well past the epoch of reionization, and that the environmental influence of a massive host, such as M31, likely extends beyond its virial radius and may have led to the quenching of Pegasus~W in the recent past.

Finally, we briefly consider the possibility that Pegasus~V and Pisces~VII could have been long-term members of the M31 system and could have been quenched by environmental processes rather than by cosmic reionization. 
Estimates for the virial radius of M31 range from around 200~kpc \citep[e.g.][]{Tamm+2012} to around 300~kpc \citep[e.g.][]{Patel+2017}. Thus, it is difficult to know if either of these UFDs are within its virial radius, but they both fall comfortably within the `backsplash' region where many dwarfs have previously been within the virial radius of their host \citep[e.g.][]{Teyssier+2012}. \citet{Buck+2019} estimate the probability that various dwarfs in the vicinity of the MW are backsplash objects by comparison to their zoom-in simulations. However, these estimates rely on both location and velocity. Although no radial velocity measurements exist for Pegasus~V and Pisces~VII, we do have another piece of pertinent information; that in order for SFHs of these UFDs to be consistent with them having been quenched by environmental effects within M31's halo, they would need to have fallen in over 10~Gyr ago. This is a strong constraint as dwarfs within the halo of a large host will gradually lose orbital energy over time due to dynamical friction. \citet{Rocha+2012} find that virtually all simulated dwarfs in a MW-like system with infall times $>$10~Gyr ago are within $\sim$100~kpc of their host at $z=0$. Thus, even if Pegasus~V and Pisces~VII turn out to be backsplash objects, it is highly unlikely that they first fell in over 10~Gyr ago and it is therefore similarly unlikely that they were quenched by environmental processes within M31's halo.

In the case of Pisces~VII there is a further possibility, that it was quenched by past interactions with M33, rather than M31. As discussed by \citet{Collins+2024}, given Pisces~VII's distance from M33 of only $\sim$100~kpc, it is well within M33's halo \citep{Patel+2018} and likely an M33 satellite. Our distance measurement for Pisces~VII is compatible with that of \citet{Collins+2024}, with ours placing it only marginally farther from both M31 and M33. Thus, we come to the same conclusion, that it is likely within M33's virial radius, but as Pisces~VII still has no radial velocity measurement, it remains unclear if it is bound to M33. Furthermore, the satellite systems of MW-mass halos are far better understood, both observationally and theoretically, than those of lower mass systems. It is presently unclear if, for example, the circumgalactic medium density in these systems is sufficient to drive quenching, let alone what their circumgalactic media were like 10~Gyr ago. However, for a satellite as low mass as Pisces~VII, this is certainly a possibility. In the near future, ongoing efforts to construct observational samples of dwarf galaxy satellite systems should provide a clearer picture of quenching and the environmental processes in these systems \citep{Hunter+2025,Li+2025}. Although, this may be a possible alternative explanation for Pisces~VII's early quenching, the simplest explanation, that is consistent with the derived SFHs, remains that both Pegasus~V and Pisces~VII were quenched by reionization.

\section{Conclusions} \label{sec:conclusions}

Using HST and VLA observations we have determined distances and SFHs for three recently discovered UFDs on the periphery of the M31 system, and placed stringent constraints on their gas contents. As found in a previous analysis \citep{McQuinn+2023}, one of the three (Pegasus~W) shows tentative signs of significant SF within the past few Gyr or less, while the others (Pegasus~V and Pisces~VII) appear to have quenched by 10~Gyr ago. The VLA \hi \ observations indicate that all three are gas-poor, with Pegasus~W having the most stringent limit ($\log M_\mathrm{HI}/M_\ast < -0.9$) despite its signs of recent SF.

The SFHs of Pegasus~V and Pisces~VII closely resemble those of low-mass MW UFDs \citep[e.g.,][]{Brown+2014,Weisz+2014,Durbin+2025} that are consistent with having been quenched by cosmic reionization.
The differing accretion histories of the MW and M31 are thought to have led to different characteristic quenching epochs for their satellites \citep[e.g.,][]{Weisz+2019, D'Souza+2021,Sacchi+2021,Savino+2025}, with many MW satellites quenching early ($\sim$10~Gyr ago) and most M31 satellites quenching more recently ($\sim$6~Gyr ago). This epoch of quenching in the MW almost coincides with that of reionization, especially given the poor temporal resolution of most ancient SFHs. Meanwhile, even UFDs previously studied in M31 follow its later quenching epoch, rather than that for the epoch of reionization \citep{Savino+2023}, calling into question whether the early quenching of MW UFDs is purely the result of cosmic reionization. Thus, our findings for Pegasus~V and Pisces~VII are an important verification that some UFDs in the M31 system did quench at ancient times, suggesting that reionization is capable of fully quenching UFDs without the assistance of environmental effects. The distinction between these two UFDs and those studied previously is likely the result of their smaller stellar (and presumably, dark matter halo) masses, with Pegasus~V and Pisces~VII falling below the mass threshold required to survive reionization quenching. However, these findings are based on relatively shallow, single-orbit HST observations, and could be considerably strengthened with deeper photometry.

Similar to \citet{McQuinn+2023}, we find evidence of an extended SFH for Pegasus~W, with SF finally ceasing less that a Gyr ago. As Pegasus~W's stellar mass appears to be close to the fiducial limit for a UFD, this may indicate that it was just able to survive reionization and continue forming stars until it was recently quenched by environmental processes in the outer halo of M31. However, our analysis suggests that this apparent recent SF in Pegasus~W's SFH is the result of two unusual populations in its CMD \citep[also identified by][]{McQuinn+2023}; a handful of stars $\sim$1~mag brighter than the HB, and an overdensity of blue straggler candidates below the HB. If Pegasus~W's CMD had neither of these then it would likely be modeled as having entirely stopped forming stars $\sim$3~Gyr ago, more in line with other similar mass UFDs in the M31 system \citep{Savino+2023}.

Ultimately, with the depth of the currently available photometry from HST it was not possible to conclusively determine the source of these stars in the CMD, either as contaminants and blue stragglers, or genuine young stars. Upcoming, deeper imaging of both Pegasus~W and Tucana~B \citep{Sand+2022} with JWST will be capable of conclusively resolving this situation and will provide new insight into the impact of reionization on UFDs and the location of the mass threshold below which they are fully quenched.

\begin{acknowledgments}
We thank the anonymous referee for comments that helped to improve this paper.
This work is based on observations made with the NASA/ESA Hubble Space Telescope, obtained at the Space Telescope Science Institute, which is operated by the Association of Universities for Research in Astronomy, Inc., under NASA contract NAS5-26555.  These observations are associated with program \# HST-GO-17316.  Support for program \# HST-GO-17316 was provided by NASA through a grant from the Space Telescope Science Institute.
This work uses VLA observations from project 23B-189. The National Radio Astronomy Observatory is a facility of the National Science Foundation operated under cooperative agreement by Associated Universities, Inc.
DJS acknowledges support from NSF grant AST-2205863.
KS acknowledges support from the Natural Sciences and Engineering Research Council of Canada (NSERC).
AK acknowledges support from NSERC, the University of Toronto Arts \& Science Postdoctoral Fellowship program, and the Dunlap Institute.
DZ acknowledges support from NSF AST-2006785 and NASA ADAP 80NSSC23K0471.
\end{acknowledgments}

%

\vspace{5mm}
\facilities{HST, VLA}


\software{\href{https://www.astropy.org/index.html}{\texttt{astropy}} \citep{astropy2013,astropy2018,astropy2022}, \href{https://photutils.readthedocs.io/en/stable/}{\texttt{photutils}} \citep{photutils}, \href{https://reproject.readthedocs.io/en/stable/}{\texttt{reproject}} \citep{reproject}, \href{https://matplotlib.org/}{\texttt{matplotlib}} \citep{matplotlib}, \href{https://numpy.org/}{\texttt{numpy}} \citep{numpy}, \href{https://scipy.org/}{\texttt{scipy}} \citep{scipy1,scipy2}, \href{https://pandas.pydata.org/}{\texttt{pandas}} \citep{pandas2},  \href{https://sites.google.com/cfa.harvard.edu/saoimageds9}{\texttt{DS9}} \citep{DS9}, \href{http://americano.dolphinsim.com/dolphot/}{\texttt{DOLPHOT}} \citep{Dolphin2000,dolphot}, \href{https://github.com/jonathansick/starfish}{\texttt{StarFISH}} \citep{StarFISH}, \href{https://dustmaps.readthedocs.io/en/latest/}{\texttt{dustmaps}} \citep{Green+2018}.}


\appendix

\section{Star formation history fitting methodology} \label{sec:SFHmethods}

Our SFH fitting methodology essentially translates \texttt{StarFISH} \citep{Harris+2001,StarFISH} from \texttt{FORTRAN} to \texttt{Python}, but also add various updates and adjusts methodology where needed, incorporating approaches from \citet{MATCH}, \citet{McQuinn+2024}, and \citet{Garling+2025}, as well as some of our own innovations.
In the following subsections we describe our CMD fitting approach and discuss how we attempt to quantify the scale of the random and systematic uncertainties in the resulting SFHs.
The key steps in the fitting process are as follows: 
1) Choosing an isochrone library or libraries. 
2) Calculating the completeness and photometric uncertainties of the real CMD. 
3) Producing a library of binned CMDs of SSPs. 
4) Fitting the maximum likelihood combination of SSPs to the real CMD. 
5) Extracting meaningful information about the SFH. 
6) Quantifying uncertainties.
7) Determining present day stellar masses and luminosities.
Each of these steps is described in detail below.

\subsection{Choice of isochrones}

There are a number of widely used stellar isochrone libraries in the literature which would suffice for generating SSPs. It is beyond the scope of this paper to review their relative strengths and weaknesses, however, it is important to select an isochrone set that spans the range of ages and metallicities relevant for the real stellar population being fit. For example, an isochrone set that does not include tracks for very young (e.g. $<$100~Myr) stars would likely be unsuitable for fitting the CMD of a young star cluster or a galaxy with an ongoing burst of SF. Equally, an isochrone set that does not compute tracks for very low metallicities (e.g. $[\mathrm{Fe/H}] < -2$) would likely not be suitable for fitting a highly metal-deficient ancient globular cluster. Beyond these basic caveats we do not have strong preferences for any particular isochrone library. However, as alluded to above (and discussed in detail in \S\ref{sec:SFH_uncertainties}) this choice likely represents the single largest source of systematic uncertainty and we therefore have chosen to compute all fits using three different isochrone sets: PARSEC \citep{Bressan+2012}, MIST \citep{Dotter+2016,Choi+2016}, and BaSTI \citep{Pietrinferni+2004,Hidalgo+2018}.

\subsection{Completeness and photometric uncertainties} \label{sec:phot_uncert}

To produce realistic mock CMDs for any SSP it is essential to first characterize the completeness and photometric uncertainties in the actual data. To do this we generate a million artificial stars with random magnitudes and colors in the ranges $20 < \mathrm{F606W} < 30$ and $-1 < \mathrm{F606W-F814W} < 2$, respectively. These are placed randomly across the real image and extracted with \texttt{DOLPHOT} in an identical manner as the real stars. The fraction of these artificial stars that are recovered (as a function of color and magnitude) allows us to determine the completeness limit of the real data, while the deviations between their input and recovered magnitudes enable the construction of a realistic error model.

To calculate the completeness and error model we first bin the artificial stars in color and magnitude. The choice of these bins is subjective, but it should be compatible with (i.e. have aligned edges and be an integer multiple in size) the bins used for the producing the SSPs (\S\ref{sec:binned_mock_CMDs}) and should extend well beyond (e.g. $\sim$1~mag) the faintest magnitude used for fitting the CMD population. For all three targets we use square bins of width 0.2~mag in the magnitude and color ranges $18 < \mathrm{F814W} < 28$ and $-0.5 < \mathrm{F606W-F814W} < 1.5$, respectively. The fraction of artificial stars recovered in each bin provides a map of completeness throughout the CMD. To calculate the error model we fit (in each bin) the 2D distribution of errors in F606W and F814W with a bivariate Gaussian distribution \citep[similar to][]{StarFISH,Garling+2025}. The parameter values of this fit are saved for every bin and allow us to add realistic photometric uncertainties to our mock SSPs.

\subsection{Binned CMDs of mock SSPs} \label{sec:binned_mock_CMDs}

Before producing mock binned CMDs, or Hess diagrams (we will use these two terms interchangeably), it is necessary to choose a grid of metallicity and age values over which SSPs will be generated. This step is subtly different for each of the three isochrone sets as the valid ranges of parameters are different for each and their pre-computed isochrones occur at different specific values. For PARSEC we use the web service\footnote{\url{stev.oapd.inaf.it/cgi-bin/cmd}} to generate an equally spaced (by 0.1~dex) set of isochrones over the metallicity range $-2.2 \le [\mathrm{M/H}] \le -1.3$, and over the logarithmic age range $6.5 \le \log t/\mathrm{yr} \le 10.1$, spaced by 0.05~dex. With this scheme the final bin essentially contain all stars formed prior (and during) the epoch of reionization. We chose to limit the maximum metallicity to $[\mathrm{M/H}] \le -1.3$ as the stellar mass of UFDs make it extremely unlikely that a more metal-rich population could exist within them, and this limit will prevent non-physical combinations of age and metallicities for these objects. We would have adopted a lower minimum metallicity than $[\mathrm{M/H}] = -2.2$ but PARSEC does not produce isochrones below this metallicity.

For MIST we used a metallicity range of $-2.5 \le [\mathrm{Fe/H}] \le -1.3$ in steps of 0.1~dex and the same age range as for PARSEC. These were generated using the MIST isochrone interpolation web service.\footnote{\url{https://waps.cfa.harvard.edu/MIST/interp\_isos.html}} Due to the more limited functionality of the BaSTI web interface we adopted a coarser binning for metallicity, covering the same range as MIST, but with 0.2~dex steps, and downloaded the pre-computed isochrones of all ages at each metallicity. From these ages we selected $\log t$ values similar to those used for PARSEC and MIST.\footnote{The full range of $\log t$ values used for BaSTI is: 7.30, 7.48, 7.60, 7.70, 7.78, 7.85, 7.90, 7.95, 8.00, 8.08, 8.15, 8.20, 8.26, 8.30, 8.38, 8.45, 8.51, 8.56, 8.60, 8.66, 8.70, 8.74, 8.81, 8.85, 8.90, 8.95, 9.00, 9.04, 9.11, 9.15, 9.20, 9.26, 9.30, 9.36, 9.40, 9.45, 9.51, 9.56, 9.60, 9.65, 9.70, 9.75, 9.80, 9.85, 9.90, 9.95, 10.00, 10.05, \& 10.10.} As we adopt a coarser age binning when interpreting the SFH (\S\ref{sec:SFH_inspection}) this slight change of $\log t$ values for BaSTI is expected to have minimal impact on the final SFH and even offers a means to verify that the final results are not overly sensitive to the binning scheme. However, we note that BaSTI does not produce isochrones for stars younger than $\sim$12.3~Myr, and thus our youngest bin is at $\log t/\mathrm{yr} = 7.30$. Given the anticipated stellar populations of the galaxies in question in this paper, this limitation is largely irrelevant. 

With the grid of age and metallicity values set for each isochrone set, the next step is to generate mock SSPs for each isochrone. To do this we adopt a Kroupa \citep{Kroupa+2001,Kroupa+2002} initial mass function (IMF), with a high-mass slope of -2.3 and a low-mass slope of -1.3, with the break occurring at 0.5~\Msol. For each isochrone we draw $10^6$ random stellar masses between 0.55~\Msol \ and the highest mass surviving point in the isochrone (with the absolute maximum stellar mass set at 100~\Msol). The lower mass limit, 0.55~\Msol, was chosen based on the photometric depth of our CMDs, such that we do not draw large numbers of stars that are many magnitude fainter than our observational completeness limit. The F606W and F814W magnitudes of each mock star are determined by interpolating along the isochrone and then shifting by a fixed distance modulus (our distance measurements; \S\ref{sec:struct_params}). Mock photometric uncertainties are added to the magnitude of all stars and they are then binned in color--magnitude space to produce a Hess diagram. These Hess diagrams use a four times finer binning scheme (i.e. 0.05~mag bins) than the binning for the completeness and photometric uncertainty model. The relationship between these two binning schemes must be an integer multiple to avoid bin edge misalignments. Stars outside the bin ranges are dropped, effectively setting a faint limit of $\mathrm{F814W}=28$ mag (see above). The completeness map is then multiplied with the mock Hess diagram of each SSP and the template is normalized with the integral of the IMF from 0.1~\Msol \ to 100~\Msol \ (relative to the fraction in the range $0.55 < M_\ast/\mathrm{M_\odot} < 100$) such that the weights of each SSP in the final fitting process corresponds to the total number of stars originally formed in that population (assuming an absolute minimum stellar mass of 0.1~\Msol).

We do not anticipate that binaries will strongly impact our mock CMDs of SSPs, as for old populations we do not reach past the MSTO, where it is more likely to find stars in binaries with comparable luminosities (under the assumption that the masses of stars in binaries are uncorrelated). However, although we do not expect binaries to significantly impact the form of SSPs CMDs, they will impact its normalization and therefore the total mass contained within the SFH fit. To include binaries we follow a similar approach to \citet{Garling+2025} and create an entirely separate set of SSP templates containing 100\% binaries. We generate another set of $10^6$ stars (as described in the previous paragraph), and then draw their companion stars from the same IMF, but this time sampling the full range down to 0.1~\Msol. The magnitudes of each of these stars and their companion are combined and then photometric uncertainties and completeness are treated as before. Each SSP binary template is then normalized in the same fashion as before, such that it corresponds to a single average star in that population. During the SFH fitting process (\S\ref{sec:SFH_maxlikefit}) the single and binary templates are averaged together, weighted according to the binary fraction, which we take to be 0.35.

We now have Hess diagrams for every SSP combination of age and metallicity for our chosen isochrone sets. Originally \texttt{StarFISH} \citep{Harris+2001} recommended that at this point the user identifies SSPs that have nearly degenerate binned CMDs and combine them if they are adjacent in age and/or metallicity. This is to prevent ambiguities in the fitted SFHs resulting from essentially random assignments of stellar mass to one or other of these degenerate populations in the fitting process. Here we instead decide to retain the original gridding of age and metallicity through the fitting process and instead eliminate these degeneracies by applying a coarser age binning scheme to the fitted SFH, as described in \S\ref{sec:SFH_inspection}.

One final component must be calculated for the library of mock CMDs, a Hess diagram of contaminants and background sources. To produce this component we take all point-like sources (following the same photometric cuts as the target CMD) in the full FoV of the image that are beyond $3r_h$ from the center of the target object. We visually inspect this to verify that structures (e.g. the RGB) in the target CMD are no longer visible. The Hess diagram of these sources is then normalized such that it sums to unity. The normalized count rates in the bins of this background binned CMD are referred to as $B_{ij}$ in the following subsection.

\subsection{Maximum likelihood combination of SSPs} \label{sec:SFH_maxlikefit}

When fitting the real Hess diagram with linear combinations of SSPs, having bins that are smaller than or comparable to the typical photometric uncertainties will not improve fit quality (but will increase computation time), while having bins that are small and sparsely populated will potentially lead to gaps or underdensities in the CMD that will decrease fit quality. However, bins that are too large will effectively smear out physically meaningful information in the CMD, also potentially degrading the fit quality. 

Similarly, while the completeness and photometric uncertainties are modeled and applied to the SSP Hess diagrams, extending the fitting range to very faint magnitudes (e.g. significantly below 50\% completeness) will stress the limits of these corrections and will not produce more reliable SFHs. Due to the steepness of the IMF, the underlying stellar population increases rapidly with decreasing luminosity, thus at faint magnitudes a small bias in the calculated completeness or photometric uncertainties could results in a large number of stars being erroneously included or excluded from the SSP Hess diagrams. Therefore, it is preferable to limit the CMD fitting region to a magnitude range where these corrections are modest.

After experimenting with various bin size choices, we chose to combine the 0.05~mag bins of the real and mock CMDs 2$\times$2 into 0.1~mag bins, as a compromise between these effects. These bins span the color range $-0.5 < \mathrm{F606W-F814W} < 1.5$ but only cover $18 < \mathrm{F814W} < 27$, ending one magnitude brighter than the faintest stars used to populate the mock CMDs. This buffer region is to alleviate edge effects such as stars scattering below the faintest magnitude cut ($\mathrm{F814W} = 28$), while none can scatter up in the opposite direction. The real CMD is also restricted to these same bins to ensure a like-for-like comparison.

During the fitting process the quantity that we wish to minimize is the difference between the bin values in the real Hess diagram and the bins of the linear combination of artificial SSPs (the weights of which constitute the SFH). However, as many of the bins contain few or no stars, performing a least squares minimization would result in a fit that is potentially significantly offset from the true maximum likelihood fit. For an individual bin in the Hess diagram the Poisson likelihood is
\begin{equation} \label{eqn:singlelike}
    l = p(n|m) = \frac{m^n e^{-m}}{n!}
\end{equation}
where $n$ is the count in the real bin and $m$ is the count rate in the mock CMD being fit. Taking the product of this quantity over all bins, the log likelihood becomes
\begin{equation} \label{eqn:loglike}
    \ln \mathcal{L} = \sum_{i,j} n_{ij} \ln m_{ij} - m_{ij} - \ln n_{ij}!
\end{equation}
where the sum goes over all combinations of $i$ and $j$ bins in the two dimensions (color and magnitude) of the binned CMD. For completeness we also define
\begin{equation} \label{eqn:weights}
    m_{ij} = w_{b}B_{ij} + \sum_{k,l} w_{kl} H_{ijkl}
\end{equation}
where $H_{ijkl}$ is the normalized $ij$ color-magnitude bin value of the mock Hess diagram for the $k$th SSP age bin and the $l$th metallicity bin (see bin definitions in \S\ref{sec:binned_mock_CMDs}), $w_{kl}$ is the weight of this SSP (i.e. the number of stars formed in this SSP), and $w_{b}$ is the number of stars in the background/contaminants component of the binned CMD. All of the SSP Hess diagrams were already generated in the previous section and 
\begin{equation}
    H_{ijkl} = (1-f_\mathrm{binary})H_{ijkl-\mathrm{single}} + f_\mathrm{binary}H_{ijkl-\mathrm{binary}}
\end{equation}
with, in this case, $f_\mathrm{binary}=0.35$. Thus, during the fitting process we simply need to find the values of $w$ that minimize $-\ln \mathcal{L}$.

As shown by \citet{MATCH} this is a relatively straightforward minimization problem that should be solvable with most established algorithms that can handle scenarios with a high number of dimensions. We use the Powell method implemented in the \texttt{scipy} package in \texttt{Python}, and have experimented with initial guesses of both a constant SFH and no past SF, to ensure a consistent solution has been reached. 

\begin{figure}
    \centering
    \includegraphics[width=0.5\columnwidth]{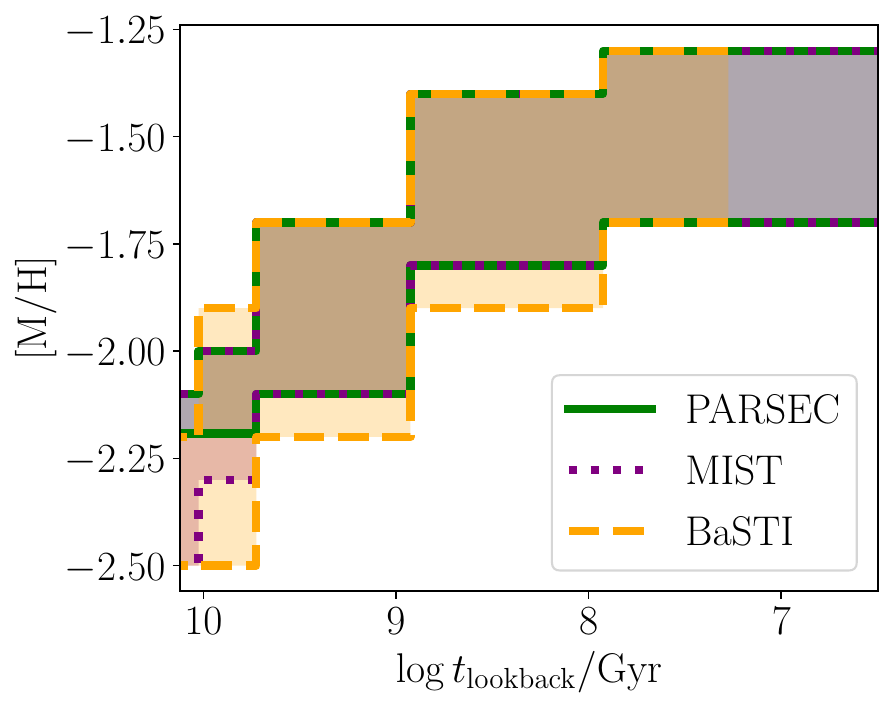}
    \caption{Permitted range of metallicities for SSPs of a given age in the mock CMD fitting process. This enforces a mild evolution in metallicity and avoids unphysical populations being included in the fit. Note that this figure indicates the allowed metallicity of populations of a given age, not the evolution of the average metallicity of the overall stellar population with age, which is of course also a function of the SFH. The blank regions of this figure effectively correspond to $w_{kl}$ values (Equation~\ref{eqn:weights}) that are fixed at zero.}
    \label{fig:Z_evo}
\end{figure}

The final consideration in the fitting process is how to implement metallicity evolution. Because of age--metallicity degeneracy in CMDs, if the metallicity is left unconstrained then unphysical stellar populations can be included, such as a metal-rich SSP formed around reionization. Thus it is desirable to restrict the range of permitted metallicities at each age. However, if this range is too restrictive then it would amount to enforcing a predetermined metallicity evolution that may or may not be appropriate for the actual data. We therefore decide to enforce only a very weak metallicity evolution by limiting what age--metallicity combinations are allowed in each age bin, but still permitting a broad range at all but the oldest ages. We prefer this approach to requiring a monotonic increase in metallicity with time \citep[as is often the case in other codes, e.g.,][]{MATCH,Garling+2025}, as although metallicity is expected to increase with time on average, it is known that galaxies (except possibly UFDs) must accrete fresh gas throughout their lives and thus the metallicity of the latest generation of stars is unlikely to be a purely monotonically increasing function with time. The disadvantage of our approach is that it is somewhat subjective and must be tailored (though mildly) to each type of galaxy being modeled. In Figure~\ref{fig:Z_evo} we show the permitted metallicity range for each isochrone set as a function of stellar population age.

\subsection{Recovery of star formation histories} \label{sec:SFH_inspection}

With the maximum likelihood weights of all of the SSPs calculated the stellar mass formed in each age bin ($M_k$) can now be trivially recovered thanks to the normalization scheme
\begin{equation} \label{eqn:SFH}
    M_k = \sum_{l} w_{kl} \times \frac{\int_{M_\mathrm{min}}^{M_\mathrm{max}} M \xi(M)\,dM}{\int_{M_\mathrm{min}}^{M_\mathrm{max}} \xi(M)\,dM}
\end{equation}
where $\xi(M)$ is the IMF and $M_\mathrm{min}$ and $M_\mathrm{max}$ are the assumed minimum and maximum masses of a star, 0.1 and 100~\Msol, respectively. This second term in Equation~\ref{eqn:SFH} is just the mean mass of a star given the IMF.

For all but the deepest and most well-populated CMDs our original age bins (\S\ref{sec:binned_mock_CMDs}) will be too fine and will produce unphysical features in the SFH. In particular, a temporal binning that is too fine tends to produce alternating bins with high and near-zero star formation rates (SFRs). This is because the mock CMDs at the two ages are nearly degenerate (given the quality of the data) and an insignificant detail can cause the fitting algorithm to give all the weight to one rather than the other. This also means that the bins in the differential SFH have a high degree of correlation with their neighboring bins, as discussed by \citet{Harris+2001} and \citet{Weisz+2011}, among others. Although cumulative SFHs are more robust to such correlated bins, the same issue can manifest as a series of plateaus and sharp increases in the cumulative stellar mass formed. Generally these problems are most severe for the oldest bins where the CMD is least constraining, but can also arise in the youngest bins where the stochastic sampling of the IMF at high stellar masses (where young populations can usually be identified) in the real stellar population, coupled with the narrow time range spanned by a single time bin, leads to poorly constrained SFRs.

To help mitigate these issues we re-bin the final SFHs that we calculate using a coarser (in time) scheme. We experimented with various binning schemes and settled on the narrowest bins that appear to largely remove the artifacts highlighted above. The bin edges in this scheme go from $\log t/\mathrm{yr} = 6.4$ to 10.0, with a constant spacing of 0.2~dex, followed by a final bin edge at 10.13 (the approximate age of the universe).

\subsection{Uncertainty estimation} \label{sec:SFH_uncertainties}

As discussed in detail by \citet{Dolphin2012,Dolphin2013}, there are both significant random and systematic uncertainties in the SFHs extracted by fitting the CMDs of observed stellar populations. Beginning with random uncertainties, we estimate the scale of uncertainties in the fitted SFH by performing a Monte Carlo resampling of the maximum likelihood mock CMD. These samples are straightforward to generate as Equation~\ref{eqn:weights} gives the count rates in every bin of the Hess diagram for the maximum likelihood SFH. Resampling is thus just a matter of sampling from a Poisson distribution in each bin. In addition we also allow the distance modulus (which was set before the CMD fitting process) to vary with a Gaussian uncertainty. For a fully robust accounting of the impact of distance uncertainty, the template Hess diagrams of every mock SSP would need to be re-computed for each Monte Carlo iteration as the locations of the stars in the CMD will shift slightly relative to the completeness and photometric uncertainty models applied. However, with space-based distance measurements from either the TRGB or the HB the uncertainty is typically smaller than the bin size (0.2~mag) used in these models, and we therefore neglect this subtlety and simply shift the SSP templates relative to the real CMD according to the new distance modulus drawn in each iteration. The fitting algorithm can then be re-run many times with these re-sampled Hess diagrams as inputs in order to characterize the random uncertainties in the fitting process. For each target we create 250 resamples in this way and calculate the 68\% confidence ranges for the SFHs.

\citet{Dolphin2013} argues that approaches of this type can lead to underestimation of the true random uncertainties, particularly for time bins with little or no SF. While we agree with this assessment, in most scenarios \citep[as also argued by][]{Dolphin2012} we expect systematic uncertainties to be the dominant source of uncertainty. We have therefore elected to follow this simpler Monte Carlo resampling method to estimate random uncertainties, rather than the Markov Chain Monte Carlo approach suggested by \citet{Dolphin2013}. We also note that resampling the best fit SFH (rather than the real data) in this way provides an internal consistency check, that is, that the fitting method returns what is put in.

To include the impact of systematic effects on our uncertainty estimates for the fitted SFHs we combine the above Monte Carlo iterations for all three isochrone libraries when calculating the 68\% confidence intervals for the final properties. The differences in these libraries reflect some of the systematic uncertainty due to incomplete knowledge of stellar evolution, however, as there are only three libraries this is likely an underestimate of the full uncertainty. In future work we plan to implement an approach similar to that of \citet{Dolphin2012}, where the input stellar isochrones can be systematically offset in magnitude and effective temperature to approximate uncertainties in stellar evolution models. 

\subsection{Determining present day stellar masses and luminosities} \label{sec:SFH_masses}

To determine the present day stellar mass of galaxies from their modeled SFHs we follow the same approach as \citet{McQuinn+2024}, adopting a recycling factor of 43\% \citep[suitable for low metallicity stellar populations and a Kroupa IMF;][]{Vincenzo+2016}. This factor represents the mass fraction of any SSP population that will eventually be returned to the ISM via stellar winds and supernovae explosions, rather than being permanently locked up in stars and stellar remnants (i.e. the stellar mass of the galaxy). This recycling is treated as instantaneous in our post hoc calculation of the stellar mass, that is, the total stellar mass accumulated over the SFH is simply multiplied by 0.57. However, as the lifetimes of massive stars are short and most of the stellar mass is contained within an ancient population, the incurred bias is negligible.

The adopted stellar mass is the median of all MC iterations for all three isochrone sets combined (\S\ref{sec:SFH_uncertainties}), and the uncertainty in this value is taken as the central 68\% confidence interval. However, this is typically quite small and neglects the important source of uncertainty stemming from the unknown binary fraction, which we set to $f_\mathrm{binary} = 0.35$. The true binary fraction in our target galaxies is unknown and, due to the depth of the photometry available, not determinable. For the same reason, the impact of binaries on the form of the template SSPs is minimal (see \S\ref{sec:binned_mock_CMDs}). Its primary effect is on the normalization of all absolute quantities in the SFH (e.g. the SFRs and the total mass formed). We therefore add (in quadrature) to the above uncertainties in the stellar mass an additional uncertainty reflecting a broad possible range of the binary fraction ($0.25 < f_\mathrm{binary} < 0.75$). This uncertainty is calculated by analytically adjusting the normalization of the SFH, assuming that binaries have zero impact on the form of the SSP Hess diagram templates. This equates to having a minimum possible uncertainty of $^{+0.139}_{-0.025}$~dex on any absolute quantity directly related to the total stellar mass formed. 

The determination of the total luminosities of the targets is more computationally involved than their stellar masses. For each of the SFHs fit to the MC realizations of the CMD we determine the total present day F606W and F814W absolute magnitude. To do this we consider each SSP included in the SFH, identifying the most massive surviving star tracked in the relevant isochrone. A mock population is then created by drawing from the IMF and converting masses to magnitudes (as in \S\ref{sec:binned_mock_CMDs}). However, unlike previously, the population drawn spans the full range of masses from the minimum stellar mass considered (0.1~\Msol) to the maximum surviving mass. The magnitudes of all of these stars are then combined and weighted according to the SFH to determine the total absolute magnitude in each band. As with the stellar masses, we adopt the median value from all the MC realizations of all three isochrone sets, and use the 68\% central confidence interval to estimate the uncertainty. As the binary fraction has a much smaller impact on the derived luminosity than on the stellar mass, we have chosen to neglect its contribution to the uncertainty. Our approach also neglects the light contributed by stellar remnants, such as white dwarfs, but this is likely to be a fairly negligible component. Finally, we note that as the SFHs were fitted to extinction-corrected CMDs, no additional correction is required at this stage, and the MC realizations already include the uncertainties stemming from the distance modulus (see \S\ref{sec:SFH_uncertainties}).

\section{Mock verifications of star formation history fits}
\label{sec:mockSFHs}

\begin{figure}
    \centering
    \includegraphics[width=0.32\linewidth]{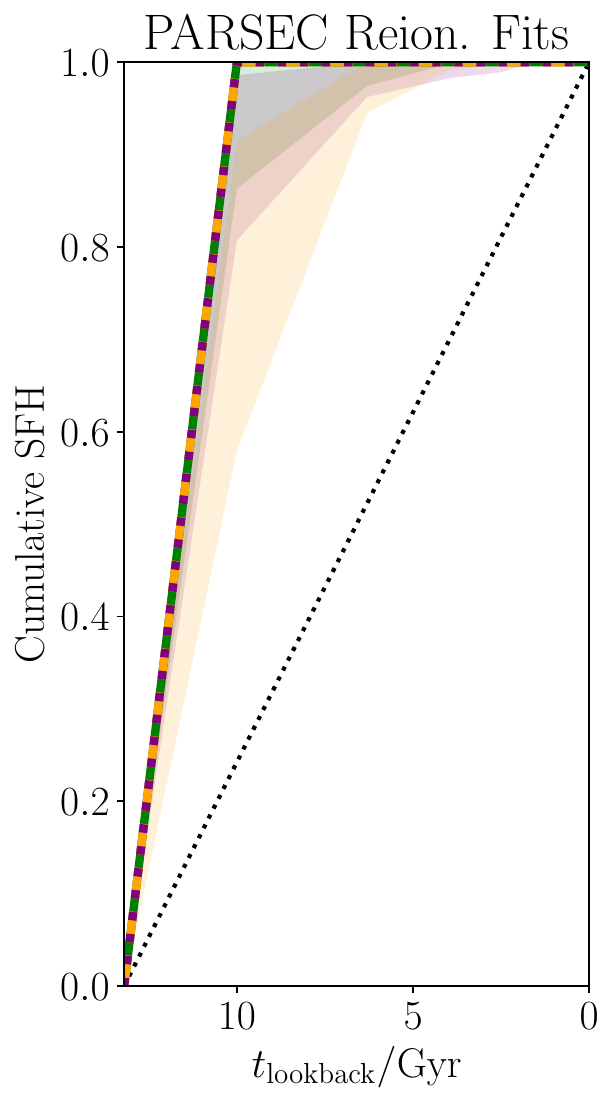}
    \includegraphics[width=0.32\linewidth]{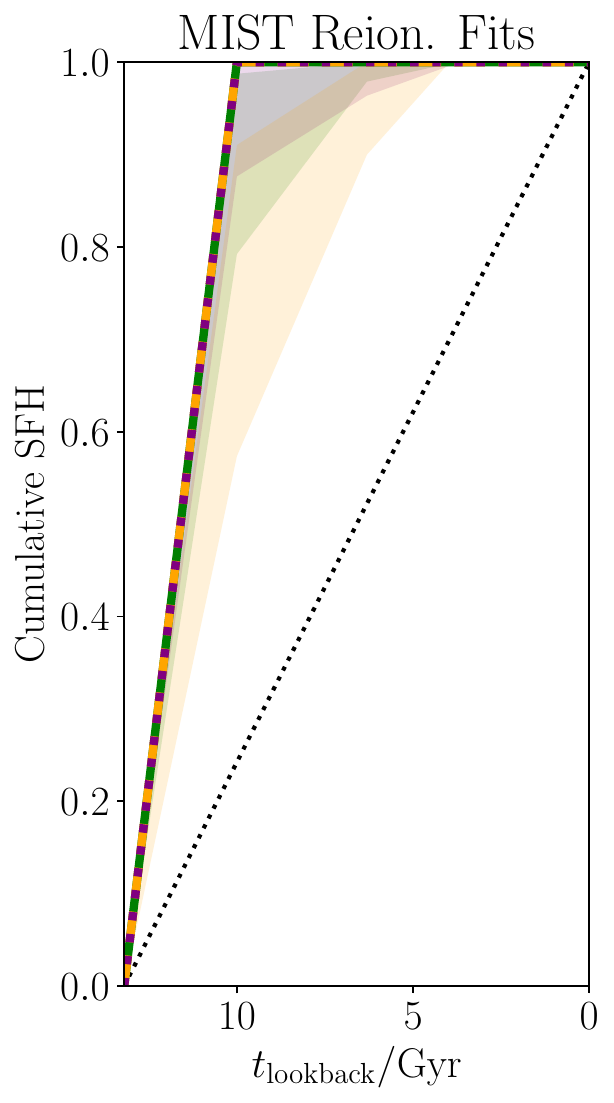}
    \includegraphics[width=0.32\linewidth]{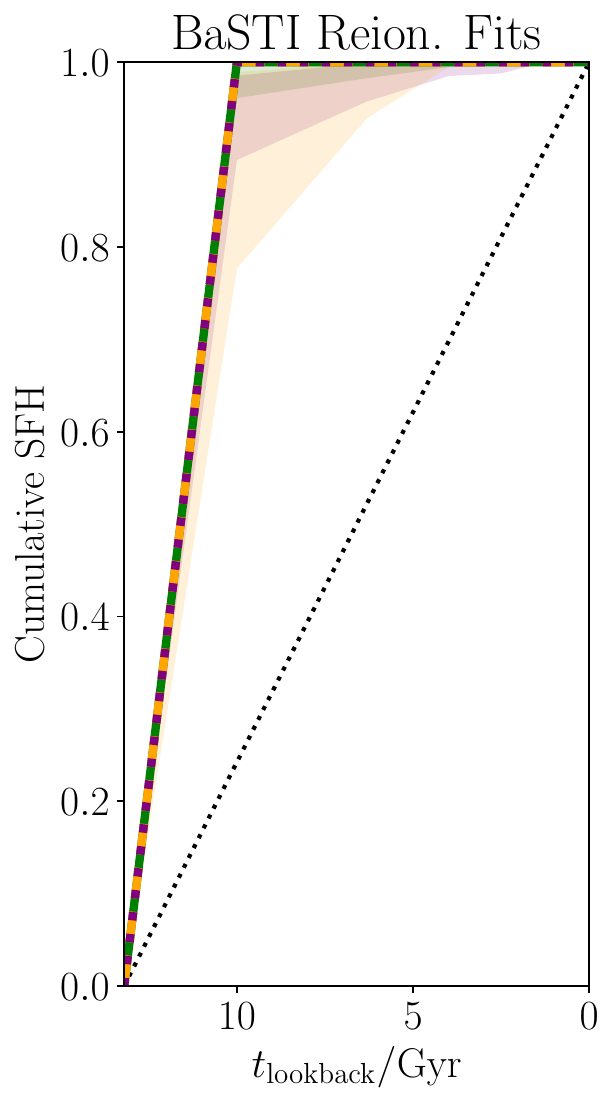}
    \includegraphics[width=0.32\linewidth]{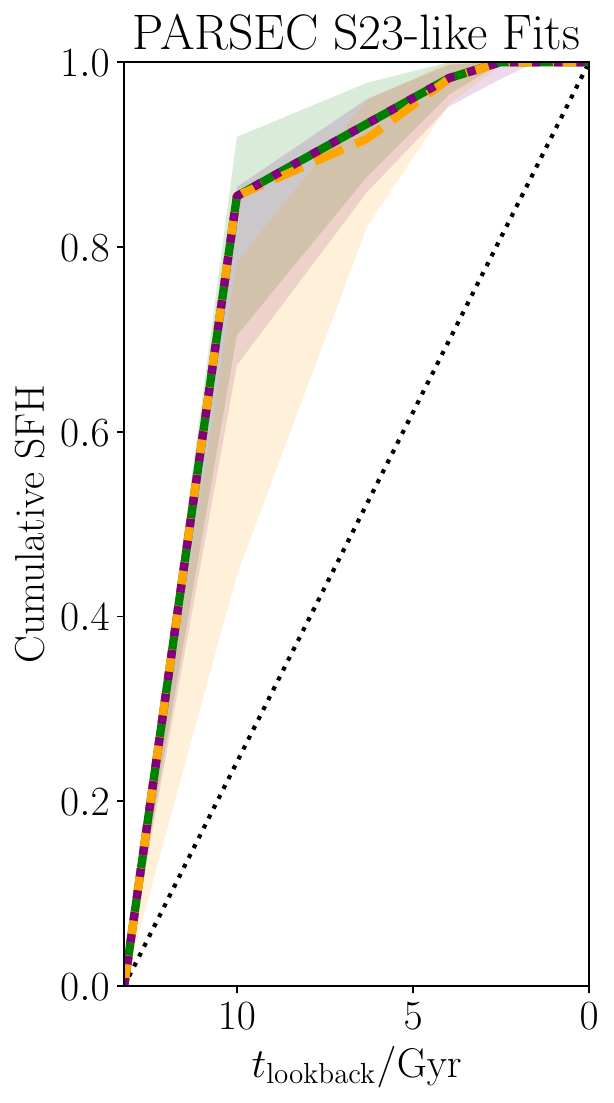}
    \includegraphics[width=0.32\linewidth]{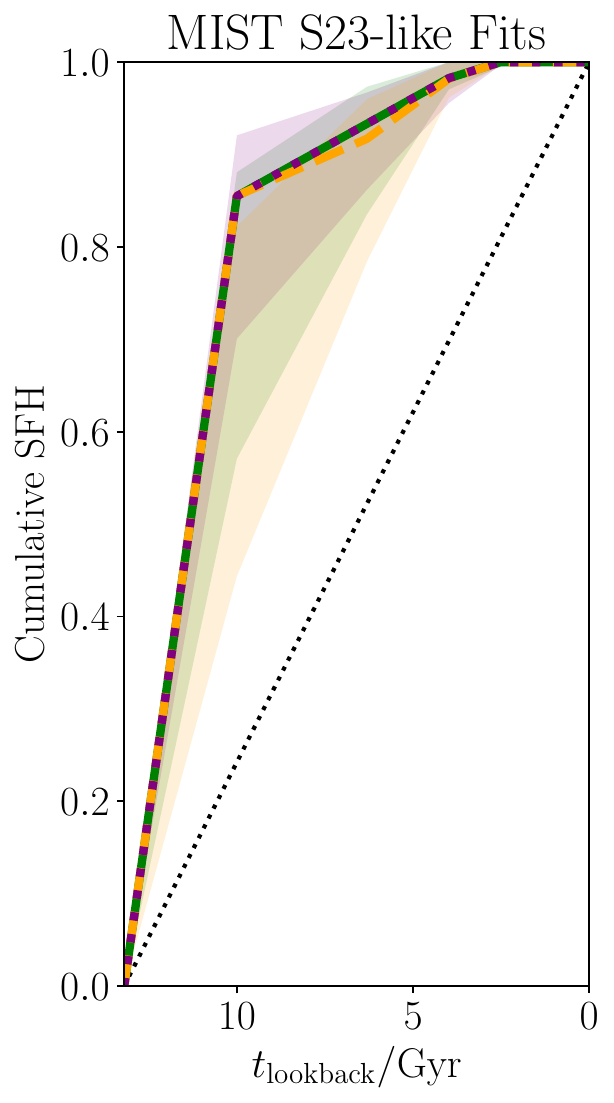}
    \includegraphics[width=0.32\linewidth]{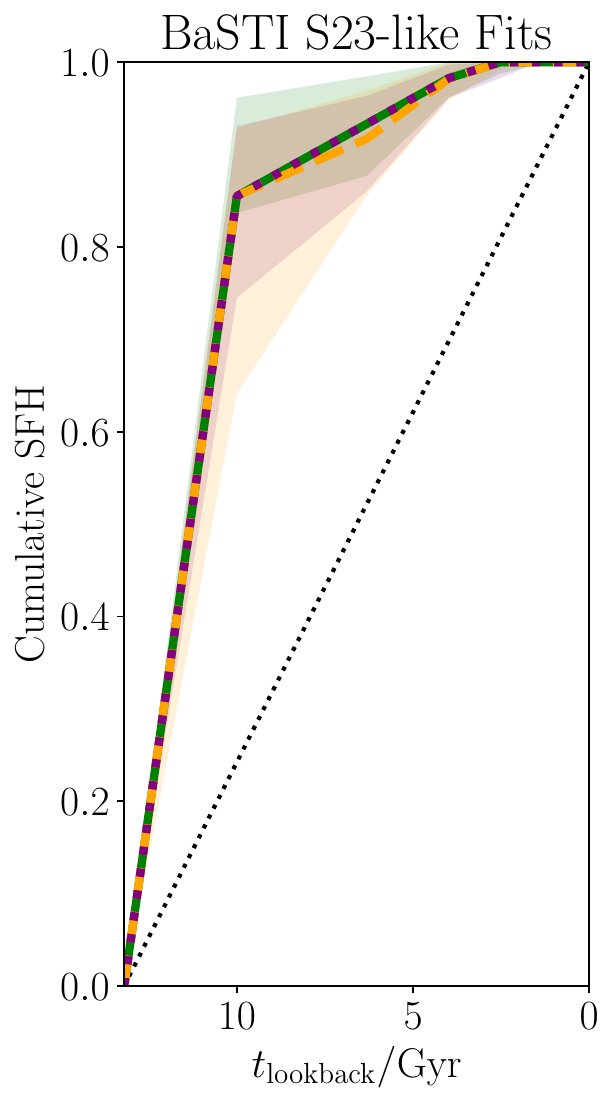}
    \caption{\textit{Top}: Model cumulative SFHs and fits for a mock dwarf entirely quenched at the epoch of reionization. \textit{Bottom}: Model cumulative SFHs and fits for a mock dwarf approximately following the evolution of the UFDs in \citet{Savino+2023}. 
    In both cases the thick lines indicate the mock SFH (line styles as in Figure~\ref{fig:cumSFHs}), while the corresponding colored bands indicate the 16th to 84th percentile range for the fits made to realizations of the CMDs from these SFHs. All the fits in the left column use PARSEC, all those in the center use MIST, and all in the right column use BaSTI.
    Note that the rise prior to reionization is in reality much steeper, but the bin reaches to 10~Gyr ago, artificially decreasing the slope of this initial build up. }
    \label{fig:mockSFHs}
\end{figure}

\begin{figure}
    \centering
    \includegraphics[width=0.32\linewidth]{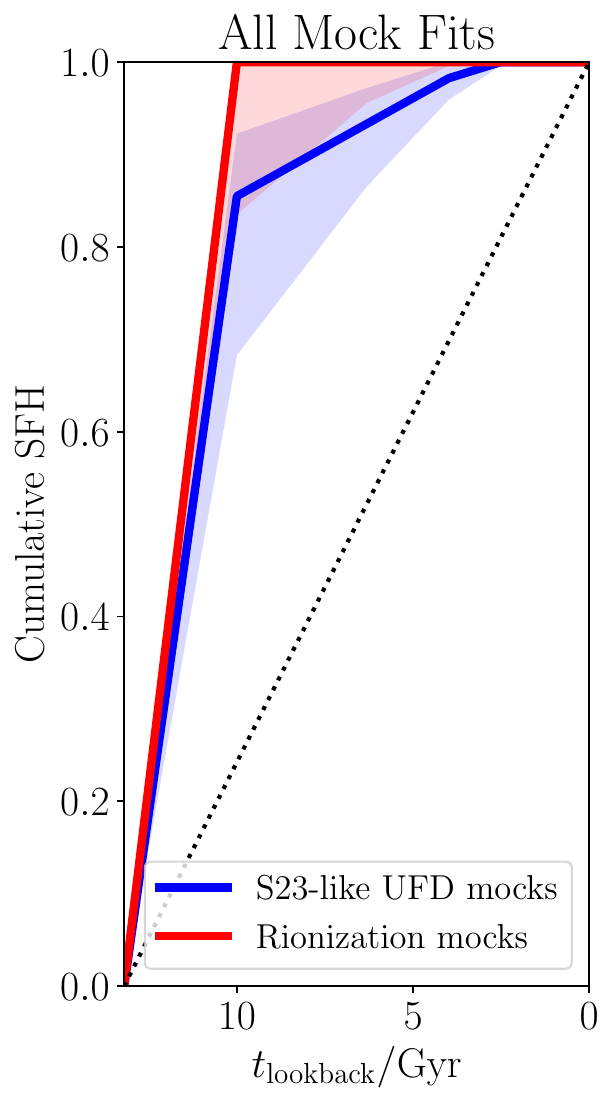}
    \caption{Cumulative SFH fits for all mocks combined for the mocks quenched by reionization (red) and those that mimic the \citet{Savino+2023} UFDs (blue). The shaded bands indicate the 16th to 84th percentiles of all fits combined for each type of mock.}
    \label{fig:allmockSFHs}
\end{figure}

To verify our SFH fitting technique we created two sets of mocks using the calculated completeness limits and photometric uncertainties for Pegasus~V, along with its distance modulus. The first set of mocks consisted of pure reionization relics with no star formation after the end of reionization. The second set were designed to approximately mimic the SFHs of other, slightly more massive M31 UFDs from \citet{Savino+2023}, forming 80\% of their stars by the end of reionization, then gradually building up the remaining 20\% through sustained (on average) star formation until about 3~Gyr ago. In both cases the present day stellar mass of the mocks was 25,000 \Msol, slightly more than Pegasus~V, but less than Pisces~VII.

The mocks were created by constructing artificial SFHs (described above) in the fine temporal resolution bins (0.05~dex), then combining the Pegasus~V template SSPs, weighted with that SFH to create a mock binned CMD. This was done for each isochrone library for both of the model SFHs, generating a total of six mocks. 

To fit the mocks we generated 100 realizations of each of their CMDs and fit each realization with each of the isochrone libraries (including the one used to produce it), exactly as for the real observations. This resulted in a total of 18 sets of 100 fits, which we used to characterize the reliability of our fits in this scenario of interest, i.e., comparing a system totally quenched by reionization to one heavily impacted by reionization, but with notable and prolonged star formation well after reionization.

Figure~\ref{fig:mockSFHs} shows the 18 sets of fits, with the top row showing mocks entirely quenched by reionization and the bottom row showing those with SFHs analogous to the UFDs in \citet{Savino+2023}. We note that although we have binned the SFHs in the same manner as those in Figure~\ref{fig:cumSFHs}, to allow direct comparison, the mocks were produced using four times higher temporal resolution (in log space). In general we see that fits appear to be quite reliable back to $\sim$7~Gyr ago, and although the errors grow at older times, the fits for the two sets of mocks appear to generally reside in different regions of the parameter space, especially if focusing primarily on PARSEC and MIST. The PARSEC and MIST fits of the BaSTI mocks (yellow bands in the left and center columns) appear to struggle most to reproduce the input SFH, although the reverse is not true; BaSTI seems able to relatively reliably fit mocks from all the isochrone sets.

In Figure~\ref{fig:allmockSFHs} we combine all of the fits for each set of mocks to assess how readily we can expect to distinguish the two scenarios. A cursory inspection of the figure reveals that there is overlap between the uncertainties of the fits to the two mock SFHs, indicating that with the currently available photometry it would not always be possible to distinguish these two scenarios. However, in this particular case almost all (5/6) of our SFH fits for Pegasus~V and Pisces~VII (Figure~\ref{fig:cumSFHs}) reside predominantly in the region above the blue uncertainty band for the \citet{Savino+2023}-like SFHs. Overall the mock fits indicate that it is more likely for fits to strongly underestimate the fraction of stars formed at early times, than it is for them to overestimate it. Thus, as our SFH fits for Pegasus~V and Pisces~VII rise so rapidly to over 90\%, this implies that those objects are indeed more consistent with an ancient quenching scenario than the SFHs of other M31 UFDs. However, deeper photometry, either from JWST or multiple-orbit observations with HST, is needed to conclusively verify this finding.

\section{TRGB location in isochrone sets}

\begin{figure}
    \centering
    \includegraphics[width=0.5\linewidth]{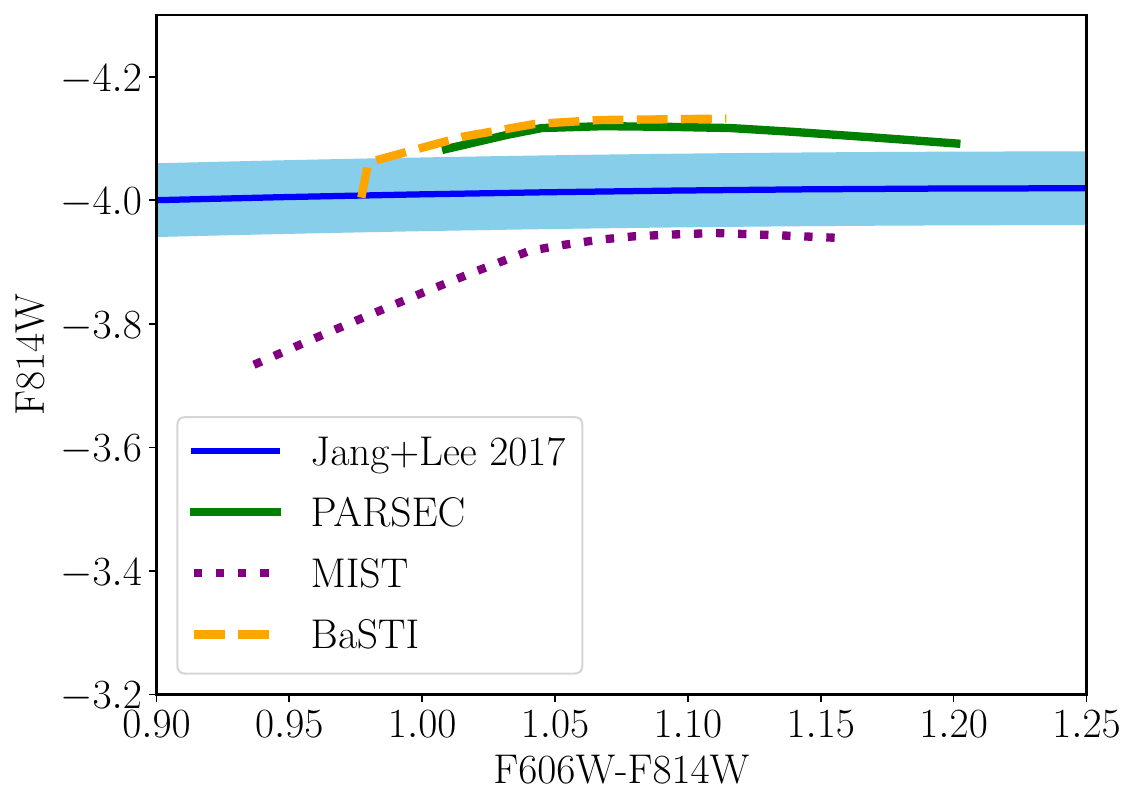}
    \caption{Absolute F814W magnitude of the TRGB as a function of F606W-F814W color (indicative of metallicity) for a $\log t/\mathrm{yr} = 10.1$ stellar population in each of the three isochrone sets that we use. The widely used emipirical TRGB calibration from \citet{Jang+2017} is shown for comparison, with the shaded band indicating their 1$\sigma$ uncertainty.}
    \label{fig:TRGBcal}
\end{figure}

In Figure~\ref{fig:TRGBcal} we show the location (in color--magnitude space) of the TRGB for each of the three isochrone sets that we use, over the metallicity range that we adopt. All values are for a population of age $\log t/\mathrm{yr} = 10.1$. Higher metal content generally produces a redder TRGB, and for colors redder than $\mathrm{F606W-F814W} = 1.05$ all of the models are within $\sim$2$\sigma$ of the \citet{Jang+2017} TRGB calibration. However, for bluer, more metal-poor populations MIST appears to deviate quite markedly from the empirical TRGB calibration. For ancient, metal-poor populations the TRGB occurs around a $\mathrm{F606W-F814W}$ color of 1.0, where the MIST TRGB is $\sim$0.2~mag fainter than the \citet{Jang+2017} value. Given that the RGBs in the galaxies studied in this work are very poorly populated, we do not make any attempt to correct for this apparent offset (with the absolute magnitude of the HB being more important). However, we note that for objects with ancient, metal-poor populations that do have well-populated RGBs, this deviation is a potential source of bias for CMD-based SFHs, particularly when using MIST isochrones. In such cases an offset to the distance modulus could be used to place the TRGB at the expected magnitude, but this would then result in other features (e.g. the HB and MSTO) being offset from their expected locations, and would also likely introduce a bias into the total stellar mass of the population fitted.


\bibliography{refs}{}
\bibliographystyle{aasjournal}



\end{document}